\documentclass
[aps,prc,fleqn,floatfix,twocolumn,superscriptaddress,twoside]{revtex4}

\usepackage[]{graphicx}

\usepackage{amsmath,amssymb,amsfonts}

\usepackage{color}
\usepackage{slashed}

\begin{document}

\title{Photon bremsstrahlung from quark jet via transverse and longitudinal scatterings: single versus multiple scatterings}

\author{Le Zhang}

\affiliation{Institute of Particle Physics and Key Laboratory of Quark and Lepton Physics (MOE), Central China Normal University, Wuhan, 430079, China  }
\affiliation{The College of Post and Telecommunication,Wuhan Institute of Technology,Wuhan 430070, China  }

\author{De-Fu Hou}

\affiliation{Institute of Particle Physics and Key Laboratory of Quark and Lepton Physics (MOE), Central China Normal University, Wuhan, 430079, China  }

\author{Guang-You Qin
}

\affiliation{Institute of Particle Physics and Key Laboratory of Quark and Lepton Physics (MOE), Central China Normal University, Wuhan, 430079, China  }

\date{\today}
\begin{abstract}

We study the production of jet-bremsstrahlung photons through the scattering with the constituents of a dense nuclear matter within the framework of deep-inelastic scattering off a large nucleus.
Applying a gradient expansion up to the second order for the exchanged three-dimensional momentum between jet and medium, we derive the single photon bremsstrahlung spectrum with the inclusion of the contributions from the transverse broadening as well as the longitudinal drag and diffusion of the hard parton's momentum.
We also compare the medium-induced photon radiation spectra for single scattering and from the resummation of multiple scatterings.
It is found that the coupling between different scatterings can give additional contribution to medium-induced photon radiation,  while for small momentum exchange, the leading contribution from the drag and diffusions to the photon emission spectra remain the same for single and multiple scatterings. 

\end{abstract}
\maketitle

\section{Introduction}

Hard partonic jets produced from early hard collisions provide very useful tools to probe the properties of hot and dense nuclear matter produced in high-energy heavy-ion collisions, such as those at the Relativistic Heavy-Ion Collider (RHIC) and the Large Hadron Collider (LHC) \cite{Wang:1991xy, Qin:2015srf, Blaizot:2015lma}.
During the propagation through the dense nuclear medium, the hard jet partons interact with the surrounding medium and tend to lose their energy and forward momentum before exiting the medium and fragmenting into hadrons.
One of the important consequences of jet-medium interaction and parton energy loss is the suppression of high transverse momentum hadron production in relativistic heavy-ion collisions as compared to that in elementary proton-proton collisions scaled by the number of binary nucleon-nucleon collisions \cite{Adcox:2001jp,Adler:2002xw,Aamodt:2010jd,CMS:2012aa,Abelev:2012hxa,Aad:2015wga}.
This phenomenon is usually denoted as jet quenching.

Parton energy loss and jet modification originate from a combination of binary elastic collisions between the hard jet partons and the medium constituents \cite{Bjorken:1982tu, Braaten:1991we, Djordjevic:2006tw, Qin:2007rn}, and medium-induced inelastic radiative/splitting processes \cite{Baier:1996kr, Baier:1996sk, Zakharov:1996fv, Gyulassy:1999zd, Gyulassy:2000fs, Gyulassy:2000er, Wiedemann:2000za, Wiedemann:2000tf, Arnold:2001ba, Arnold:2002ja, Guo:2000nz, Wang:2001ifa, Majumder:2009ge}.
For light flavor jet quenching, especially the suppression of single inclusive hadron production at high transverse momentum, the medium-induced gluon radiation is usually regarded as the most important parton energy loss mechanism \cite{Wicks:2005gt,Qin:2007rn,Schenke:2009ik}, while for the nuclear modification of heavy quarks and heavy flavor mesons with large finite masses, elastic processes may assume more important contribution, especially at low and intermediate transverse momentum regime \cite{Moore:2004tg, Mustafa:2004dr, Wicks:2005gt, Qin:2009gw, Cao:2013ita, Cao:2015hia}.
Various theoretical formalisms of jet quenching have been developed to study the collisional and radiative parts of the energy loss experienced by the hard jet partons when they traverse the dense nuclear medium (see Ref. \cite{Armesto:2011ht} and references therein for a more detailed comparison of the different jet quenching formalisms). 
Based on these theoretical formalisms, sophisticated phenomenological studies have also been performed to investigate the manifestation of jet-medium interaction in different final state observables, such as the suppression of single inclusive high hadron production at high transverse momentum regime \cite{Bass:2008rv, Armesto:2009zi, Chen:2010te}, the nuclear modification of dihadron and photon-hadron pair correlations \cite{Zhang:2007ja, Majumder:2004pt, Qin:2009bk, Renk:2008xq}, etc.
One of the main purposes of various jet quenching studies is to extract the values of jet transport coefficients, such as the transverse momentum broadening rate $\hat{q} = d\langle p_{T}^2 \rangle /dt $ and the longitudinal momentum loss rate $\hat{e} = d\langle E \rangle /dt$, in dense nuclear media via the systematic comparison with the available jet modification data (e.g., see Ref. \cite{Burke:2013yra}).

Currently, most calculations of radiative energy loss mainly focus on the gluon radiation induced by transverse momentum broadening experienced by the jet parton when propagating through the dense nuclear medium.
When a jet parton scatters off the medium constituents, both transverse and longitudinal momenta are exchanged between the propagating jet and the traversed medium \cite{Majumder:2008zg, Qin:2012fua, Abir:2014sxa}.
The longitudinal momentum loss experienced by the propagating jet partons has been studied in many jet quenching calculations, but only in the context of evaluating purely collisional energy loss either by the leading hard parton \cite{Wicks:2005gt, Djordjevic:2006tw, Qin:2007rn, Schenke:2009ik}, or by the radiated partons when studying the energy loss and the medium modification of full jet showers \cite{Qin:2009uh, Neufeld:2009ep, Qin:2010mn, Qin:2012gp, Chang:2016gjp}.
Recently, the influence of longitudinal momentum exchange between the jet parton and the medium constituents on the stimulated radiation vertex has been investigate in Ref. \cite{Qin:2014mya, Abir:2015hta} by allowing the exchanged longitudinal momentum to be of the order of the transverse momentum.

In this work we study the real photon radiation from a jet parton which propagates through a dense nuclear medium and experiences both transverse and longitudinal scattering off the gluon field of the medium.
This work serves as an intermediate step to evaluate medium-induced gluon emission from a hard parton interacting with the dense nuclear medium.
It is also an extension of the previous work performed by one of the authors in Ref. \cite{Qin:2014mya} in which the leading contribution to parton-photon double differential spectrum has been obtained in the presence of both transverse and longitudinal momentum exchanges.
Here, we attempt to derive the single photon radiation spectrum induced by jet-medium interaction by integrating out the momentum of the final outgoing parton.
As will be shown, in order to obtain the leading contribution from longitudinal and transverse momentum diffusion to the single photon radiation spectrum, one needs to proceed one step further in the momentum gradient expansion as compared to the previous work in Ref. \cite{Qin:2014mya}.
The single photon emission spectrum obtained in this work can be directly used as the input to evaluate the production of jet-medium photons in relativistic heavy-ion collisions, which are expected to provide significant contribution to the direct photon production at the intermediate transverse momentum regime \cite{Fries:2002kt, Qin:2009bk}.
As an interesting check, we also compare the single photon bremsstrahlung spectrum from the resummation of multiple scattering scenario and that from single scattering calculation. 
We find that the coupling between different scatterings experienced by the hard propagating parton may give additional contribution to the medium-induced photon emission, while for small momentum exchange and only considering the leading contribution from the drag and the diffusions to the single photon emission spectrum, the results from single and multiple scattering scenarios remain the same. 

The paper is organized as follows.
In Sec. II, we provide a brief introduction to the photon bremsstrahlung process from a hard parton produced in deep-inelastic scattering (DIS) off a large nucleus at leading-twist level.
In Sec. III, we investigate the induced photon radiation when a hard parton experiences a single scattering in the dense nuclear medium (at twist-four level).
Applying a gradient expansion for the exchanged three-dimensional momentum between the jet parton and the nuclear medium, the contributions from the transverse momentum diffusion as well as the longitudinal momentum drag and diffusion to the single photon emission spectrum are computed.
In Sec. IV, we present the computation of the single photon bremsstrahlung spectrum from a hard parton which experiences multiple scatterings in the dense nuclear medium (i.e., by resumming all-twist contributions).
By resumming the number of multiple scatterings experienced by the propagating hard parton, we compute the single photon spectrum at all-twist level, and compare to the twist-four result obtained in Sec. III.
Sec. V contains our summary.

\section{Photon bremsstrahlung in DIS at leading twist}

For completeness, in this section we present a short introduction to the photon bremsstrahlung process from a jet parton produced in semi-inclusive process of deep-inelastic scattering (DIS) off a large nucleus at leading twist.
We consider the following process:
\begin{eqnarray}
e(L_1) + A(P_A) \to e(L_2) + q(l_q) + \gamma(l) + X,
\end{eqnarray}
where $L_1$ and $L_2$ denote the momenta of the incoming and outgoing leptons, $l_q$ and $l$ are the momenta of the produced hard quark and the bremsstrahlung photon, and $P_A= Ap$ is the momentum of the incoming nucleus with the atomic number $A$ (i.e., each nucleon in the nucleus carries the same momentum $p$).
In this work, we utilize the light-cone notation for four-momentum vectors, e.g., $p = [p^+, p^-, \mathbf{p}_\perp]$, with
\begin{eqnarray}
p^+ = \frac{1}{\sqrt{2}}(E+p_z), p^- = \frac{1}{\sqrt{2}}(E-p_z).
\end{eqnarray}
We note that in the Briet frame, the virtual photon $\gamma^*$ carries a four-momentum $q = L_2 - L_1 = [-x_Bp^+, q^-, \mathbf{0}_\perp]$ and the nucleus has a four-momentum $P_A = Ap = A[p^+, 0, \mathbf{0}_\perp]$, where $x_B$ is the Bjorken variable defined as: $x_B = Q^2 / (2p^+q^-)$, with $Q^2=-q^2$ the invariant mass of the virtual photon.

The differential cross section for the lepton production from the above semi-inclusive DIS process can be written as follows:
\begin{eqnarray}
E_{L_2} \frac{d\sigma}{d^3\mathbf{L}_2} = \frac{\alpha_{e}}{2\pi s} \frac{1}{Q^4} L_{\mu\nu} {W^{\mu\nu}},
\end{eqnarray}
where $\alpha_e$ is the electromagnetic coupling and $s=(p+L_1)^2$ is the invariant mass of the lepton-nucleon collision system.
The leptonic tensor $L_{\mu\nu}$ is given by:
\begin{eqnarray}
L_{\mu\nu} = \frac{1} {2}{\rm Tr}[\slashed{L}_1 \gamma_\mu \slashed{L}_2 \gamma_\nu].
\end{eqnarray}
The hadronic tensor $W^{\mu\nu}$ is defined as:
\begin{eqnarray}
W^{\mu\nu} \!\!&=&\!\! \sum_{X} (2\pi)^4 \delta^4 (q + P_A - P_X) \nonumber\\ \!\!&\times&\!\!
\langle A| J^\mu(0)|X\rangle \langle X|J^\nu(0)|A\rangle.
\end{eqnarray}
In the above expression, $|A\rangle$ represents the initial state of an incoming nucleus $A$, $|X\rangle$ denotes the final hadronic (or partonic) states, and the sum over $X$ runs over all possible final states except the outgoing hard quark jet and the radiated real photon.
The electromagnetic current $J^\mu = Q_q \psi_q \gamma^\mu \psi$ is for a quark of flavor $q$ and the electric charge $Q_q$ (in unit of the positron charge $e$).
The hadronic tensor contains the detailed information about the final state interaction between the stuck quark and the traversed dense nuclear medium, and is the main focus of our current work.

\begin{figure}[thb]
\includegraphics[width=1.0\linewidth]{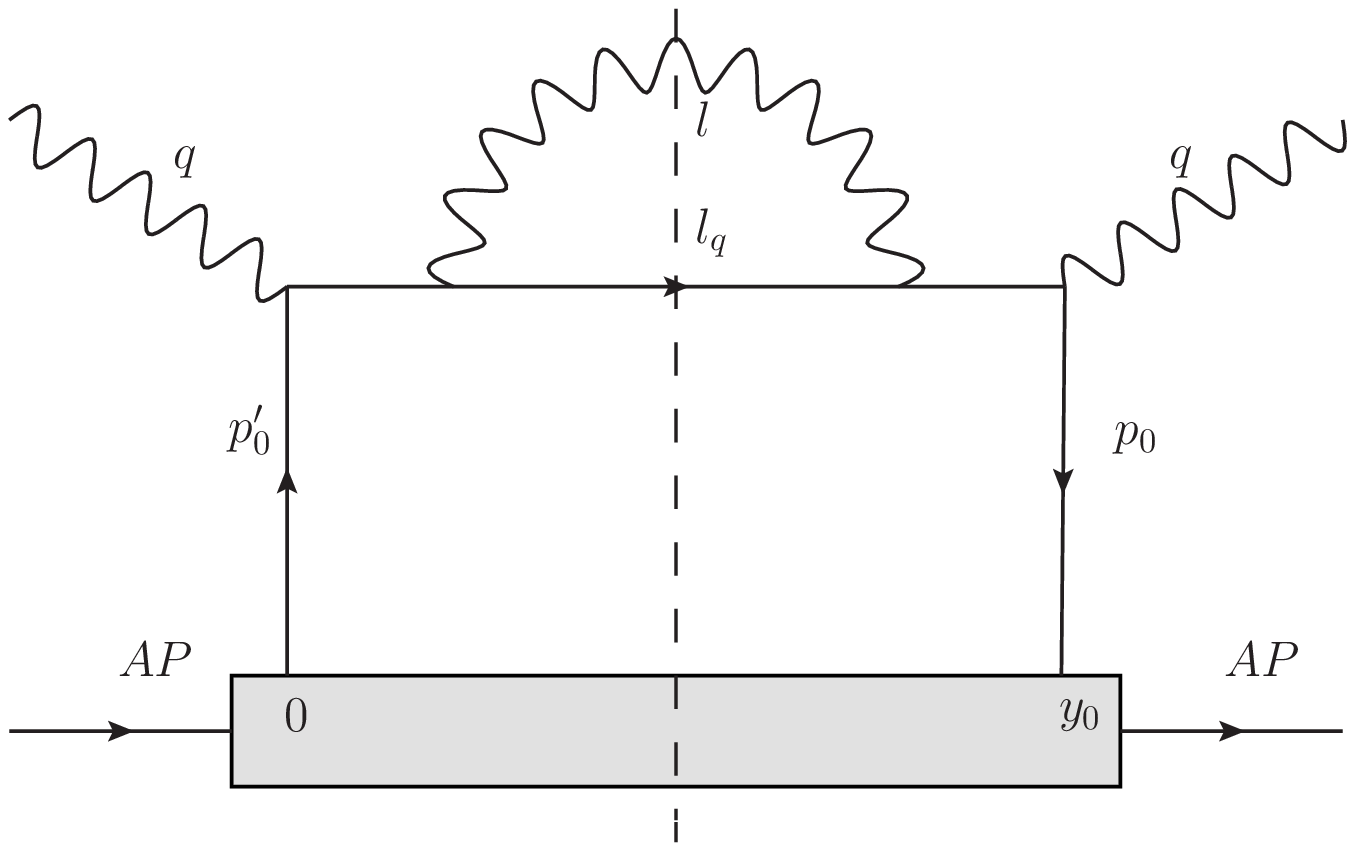}
 \caption{Leading twist contribution to the hadronic tensor.
} \label{leadingtwist}
\end{figure}

In Fig. \ref{leadingtwist} we show the photon bremsstrahlung process in the semi-inclusive DIS at leading twist.
In this work, we use the light-cone gauge ($A^-=0$), and other types of diagrams are power-suppressed.
Fig. \ref{leadingtwist} represents the process that a hard quark is first excited by the virtual photon from one nucleon of the nucleus, then radiates a real hard photon and exits without further interaction with the nuclear medium.
One may first write down the expression of the hadronic tensor for Fig. \ref{leadingtwist} as follows:
\begin{eqnarray}
W_0^{A \mu\nu}  \!\!&=&\!\!  \sum_q Q_q^4 e^2 \int \frac{d^4l}{(2\pi)^4} (2\pi)\delta(l^2) \int \frac{d^4l_q}{(2\pi)^4} (2\pi) \delta(l_q^2) \nonumber\\\!\!&\times&\!\!
\int d^4y_0 e^{iq\cdot y_0}  \int d^4z \int d^4z' \int \frac{d^4 q_1}{(2\pi)^4} \int \frac{d^4 q_1'}{(2\pi)^4} \nonumber\\\!\!&\times&\!\!
e^{-i q_1 \cdot (y_0-z)} e^{-i q_1' \cdot (z'-y_0')} e^{-il_q\cdot (z-z')} e^{-i l\cdot (z-z')}  \nonumber\\\!\!&\times&\!\!
\langle A |\bar{\psi}(y_0) \gamma^\mu \frac{\slashed{q}_1}{q_1^2 - i\epsilon} \gamma^\alpha \slashed{l}_q \gamma^\beta \frac{\slashed{q}_1'}{q_1'^2 + i\epsilon} \gamma^\nu \psi(0)|A\rangle \nonumber\\\!\!&\times&\!\!
G_{\alpha\beta}(l).
\end{eqnarray}
Here $G_{\alpha\beta}(l)$ is the sum of the photon's polarizations:
\begin{eqnarray}
G_{\alpha\beta}(l) = -g_{\alpha \beta} + \frac{n_\alpha l_\beta + n_\beta l_\alpha}{n\cdot l},
\end{eqnarray}
where $n=[1,0,\mathbf{0}_\perp]$ is the light-cone vector ($n\cdot l = l^-$).
To simplify the above expression, we may first integrate over the photon radiation locations $z$, $z'$ and obtain two momentum conservation $\delta$ functions.
Then we may perform the integration over the four-momenta $q_1$, $q_1'$, and get the relation $q_1 = q_1' = p_0 + q$.
By re-introducing the four-momentum $p_0 = l_q + l - q$, the hadronic tensor can be written as follows:
\begin{eqnarray}
W_0^{A \mu\nu} \!\!&=&\!\! \sum_q Q_q^4 e^2 \int \frac{d^4l}{(2\pi)^4} (2\pi)\delta(l^2) \int \frac{d^4l_q}{(2\pi)^4} (2\pi) \delta(l_q^2)\nonumber\\\!\!&\times&\!\!
\int \frac{d^4p_0}{(2\pi)^4} (2\pi)^4 \delta^4(q + p_0 - l_q -l) \int d^4y_0 e^{-ip_0\cdot y_0} \nonumber\\\!\!&\times&\!\!
\langle A |\bar{\psi}(y_0) \gamma^\mu \frac{\slashed{q}_1}{q_1^2 - i\epsilon} \gamma^\alpha \slashed{l}_q \gamma^\beta \frac{\slashed{q}_1'}{q_1'^2 + i\epsilon} \gamma^\nu \psi(0)|A\rangle \hspace{-12pt}\nonumber\\\!\!&\times&\!\!
G_{\alpha\beta}(l).
\end{eqnarray}
Considering the limit of very high energy and collinear radiation, one can neglect the $\perp$-component of quark field operators and factor out the one-nucleon state from nucleus state as follows:
\begin{eqnarray}
\!\!&&\!\! \langle A| \bar{\psi}(y_0) \hat{O} \psi(0) |A\rangle = A C_p^A \langle p| \bar{\psi}(y_0) \hat{O} \psi(0) |p\rangle
\nonumber\\\!\!&&\!\!
\approx A C_p^A \langle p| \bar{\psi}(y_0) \frac{\gamma^+}{2} \psi(0) |p\rangle {\rm Tr}[\frac{\gamma^-}{2} \hat{O}],
\end{eqnarray}
where $C_p^A$ denotes the probability of finding a nucleon state with a momentum $p$ inside the nucleus $A$.
$\gamma^+$ and $\gamma^-$ are defined as:
\begin{eqnarray}
\gamma^+ =  \frac{1}{\sqrt{2}}(\gamma^0 + \gamma^3), \gamma^- = \frac{1}{\sqrt{2}}(\gamma^0 - \gamma^3).
\end{eqnarray}
Now we can first integrate out the four-momentum $l_q$, and then use the photon on-shell condition $\delta(l^2)$ to integrate over $l^+$, i.e., $l^+ = l_\perp^2 / (2l^-)$.
The final quark's on-shell condition gives,
\begin{eqnarray}
(2\pi) \delta(l_q^2) = \frac{2\pi}{2 p^+q^-(1-y)} \delta(x_0 - x_B - x_L),
\end{eqnarray}
where $y=l^-/q^-$ is the fraction of the forward momentum carried by the radiated photon from the parent quark, and the momentum fraction $x_L = l_\perp^2/[2p^+q^-y(1-y)]$ is related to the photon formation time as: $\tau_{\rm form} = 1/(x_Lp^+)$.
Using the final quark's on-shell condition, one may now integrate out the momentum $p_0^+ = p^+ x_0$.
The integration over $y_0^+$ and $\mathbf{y}_{0\perp}$ gives a three-$\delta$-function, which can be used to integrate over $p_0^-$, $\mathbf{p}_{0\perp}$ and sets $p_0^- \to 0$ and $\mathbf{p}_{0\perp} \to 0$.
With the above simplifications, the hadronic tensor now reads:
\begin{eqnarray}
W_0^{A \mu\nu} \!\!&=&\!\! A C_p^A \sum_q Q_q^4 \frac{\alpha_e}{2\pi} \int dy \int \frac{d^2l_\perp}{\pi l_\perp^2} \frac{(2\pi)f_q(x_B+x_L)}{8p^+(q^-)^2x_L} \nonumber\\\!\!&\times&\!\!
 {\rm Tr} [ \frac{\gamma^-}{2} \gamma^\mu \slashed{q}_1 \gamma^\alpha \slashed{l}_q \gamma^\beta \slashed{q}_1'\gamma^\nu ] G_{\alpha\beta}(l), \ \ \ \
\end{eqnarray}
where $f_q(x)$ is the quark parton distribution function, with $x$ the fraction of the forward momentum carried by the quark from the nucleon. It is given as:
\begin{eqnarray}
f_q(x) = \int \frac{dy_0^-}{2\pi} e^{-ixp^+y_0^-} \langle p| \bar{\psi}(y_0^-) \frac{\gamma^+}{2} \psi(0)|p \rangle.
\end{eqnarray}
One may carry out the trace using the commutation relations of $\gamma$ matrices: $\{\gamma^+, \gamma^-\} = 2$, $\{\gamma^\pm, \gamma^\pm\} = 0$, and $\{\gamma^\pm, \gamma_\perp\} = 0$.
The final expression for the hadronic tensor at leading twist takes the following form:
\begin{eqnarray}
\frac{dW_0^{A\mu\nu}}{dl_\perp^2 dy} = A C_p^A \sum_q Q_q^4  \frac{\alpha_e}{2 \pi} \frac{P(y)}{l_\perp^2} (-g_\perp^{\mu\nu})  (2\pi) f_q(x_B+x_L), \ \ \ \
\end{eqnarray}
where $P(y)=[1+(1-y)^2]/y$ is the quark-to-photon splitting function, and $g_\perp^{\mu\nu} = g^{\mu\nu} - g^{\mu-}g^{\nu+} - g^{\mu+} g^{\nu-}$.
In the following, we investigate the effect of single and multiple scatterings experienced by the propagating hard quark on the photon bremsstrahlung process.

\section{Photon bremsstrahlung from single scattering}

In this section, we present the computation of medium-induced single photon radiation spectrum from a hard quark undergoing a single scattering from the dense nuclear medium. This corresponds to the contribution involving twist-four parton distribution in the nucleus as compared to the leading twist contribution presented in the previous section.
Fig.~\ref{11twist} shows the central-cut diagrams at the twist-four level, i.e., diagrams with a single scattering in both the amplitude and the complex conjugate.
The non-central-cut diagrams at the twist-four level, i.e., two gluon insertions in the amplitude and zero insertion in the conjugate (or vice versa), are shown in Fig.~\ref{20twist}.
One may write down the expressions for the hadronic tensors for all the diagrams listed in Fig. \ref{11twist} and Fig. \ref{20twist}, and compute their contributions to the single photon radiation spectrum.
Here we only provide the details for the calculation of Fig. \ref{11twist}(a).
The calculations for other diagrams are analogous and their main results are provided in the Appendix.

Our goal is to derive a closed expression for single photon radiation spectrum by a quark jet when traversing and interacting with a dense nuclear matter.
For this purpose, we use the following power counting for the exchanged gluons between the hard quark and the medium constituents.
We use $Q$ for the hardest momentum scale and $\lambda$ for a small dimensionless parameter, i.e., $\lambda Q$ denotes a softer momentum scale.
If one considers a nearly on-shell projectile parton with a momentum $q = (q^+, q^-, {q}_\perp) \sim (\lambda^2 Q, Q, 0)$ scattering off a nearly on-shell target parton traveling in the opposite direction with a momentum $\sim (Q, \lambda^2 Q, 0)$, then the exchanged gluons will be the standard Glauber gluons, i.e., the gluons carry momenta $\sim (\lambda^{2} Q, \lambda^{2} Q, \lambda Q)$ \cite{Idilbi:2008vm}.
If one allows the target parton not to be on shell, then the exchanged gluons can carry a longitudinal momentum component of the order of the transverse components, i.e., $\sim (\lambda^{2} Q, \lambda Q, \lambda Q)$; one sometimes refers to such gluons as the {longitudinal}-Glauber gluons \cite{Qin:2012fua}. 
We will investigate the effect of transverse and longitudinal momentum transfers (of the same order) on the photon radiation spectrum from a hard quark which interacts with the medium constituent via either single (this section) or multiple (next section) scatterings.

Fig. \ref{11twist}(a) shows the process that a hard virtual photon strikes a quark in the nucleus with momentum $p_0'$ ($p_0$ in the complex conjugate) at the location $y_0'=0$ ($y_0$ in the complex conjugate).
The struck quark is then sent back to the nucleus, carrying momentum $q_1'$ ($q_1$ in the complex conjugate).
During its propagating through the nucleus, the hard quark scatters off the gluon field at the location $y_1'$ ($y_1$ in the complex conjugate), and picks up a momentum $p_1'$ ($p_1$ in the complex conjugate). We denote the radiated photon's momentum as $l$ and the final outgoing quark's momentum as $l_q$.
The contribution to the hadronic tensor from Fig. \ref{11twist}(a) is given by:
\begin{widetext}
\begin{eqnarray}
\label{bb}
W^{A\mu\nu}_{(a)} \!\!&=&\!\! \sum_q Q_q^4 e^2 g^2 \frac{1}{N_c} {\rm Tr}\left[T^{a_1} T^{a_1'} \right]
\int \frac{d^4l}{(2\pi)^4} (2\pi)\delta(l^2) \int \frac{d^4l_q}{(2\pi)^4} (2\pi)\delta(l_q^2) \int d^4y_0 e^{iq\cdot y_0} \int d^4y_1\int d^4y_1'\int d^4z \int d^4z'
\nonumber\\\!\!&\times&\!\!
  \int \frac{d^4q_1}{(2\pi)^4} e^{-iq_1\cdot (y_0-z)} \int \frac{d^4\bar{q}_1}{(2\pi)^4} e^{-i\bar{q}_1\cdot (z-y_1)}e^{-il\cdot (z-z')} e^{-il_q\cdot (y_1-y_1')}\int \frac{d^4\bar{q}_1'}{(2\pi)^4} e^{-i\bar{q}_1'\cdot (y_1'-z')}\int \frac{d^4q_1'}{(2\pi)^4} e^{-iq_1'\cdot (z'-y_0')}
\nonumber \\\!\!&\times&\!\!
\langle A | \bar{\psi}(y_0) \gamma^\mu \frac{\slashed{q}_1}{q_1^2 - i\epsilon}\gamma^\alpha  \frac{\slashed{\bar{q}}_1}{\bar{q}_1^2 - i\epsilon} \slashed{A}^{a_1}(y_1)\slashed{l}_q
\slashed{A}^{a'_1}(y'_1) \frac{\slashed{\bar{q}}_1'}{\bar{q}_1'^{2} + i \epsilon}
 \gamma^\beta \frac{ \slashed{q}_1'}{{q}_{1}'^2 + i \epsilon}
 \gamma^\nu \psi(0) |A\rangle 
G_{\alpha\beta}(l) .
\end{eqnarray}

\begin{figure}[thb]
\includegraphics[width=1.0\textwidth]{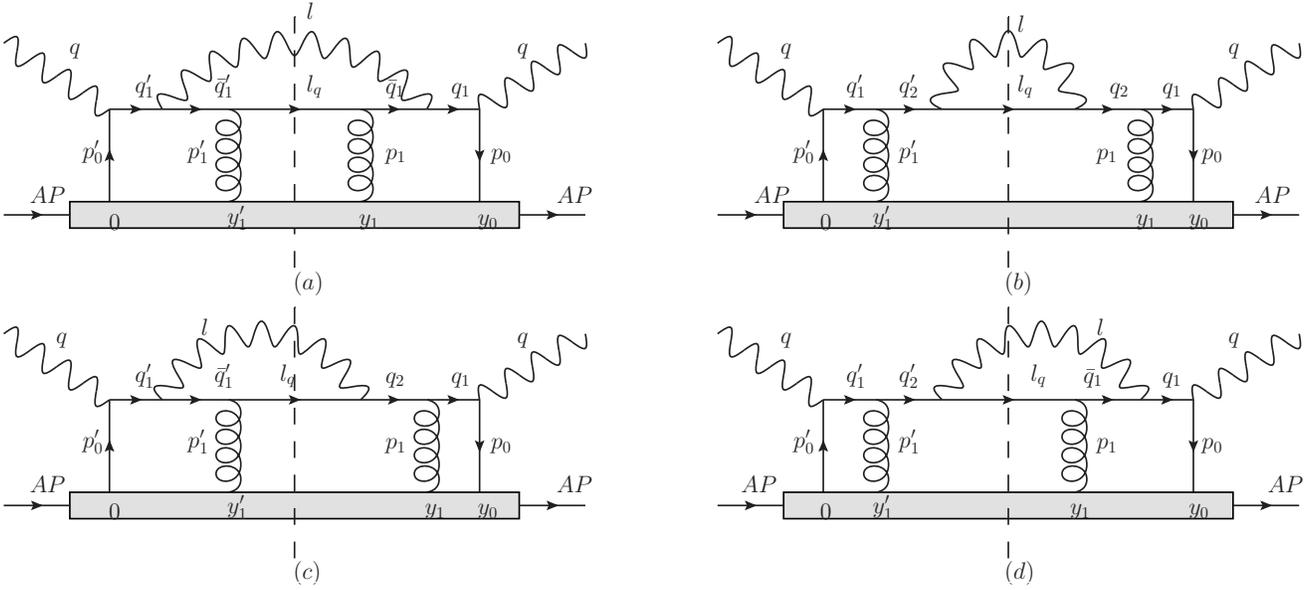}
\caption{Central-cut diagrams: a single scattering with the medium in both the amplitude and the complex conjugate. 
}
\label{11twist}
\end{figure}
\begin{figure}[thb]
\includegraphics[width=1.0\textwidth]{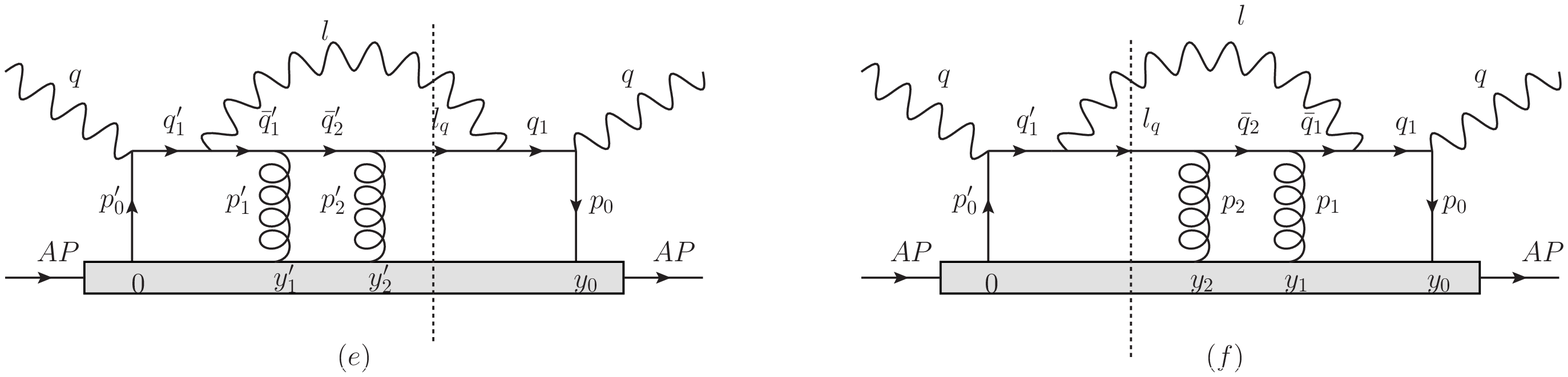}
\includegraphics[width=1.0\textwidth]{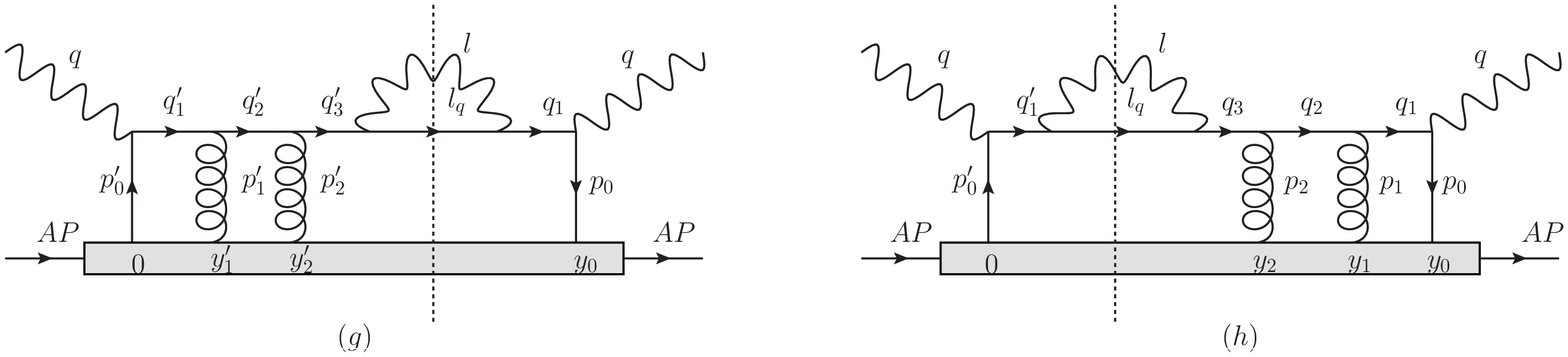}
\includegraphics[width=1.0\textwidth]{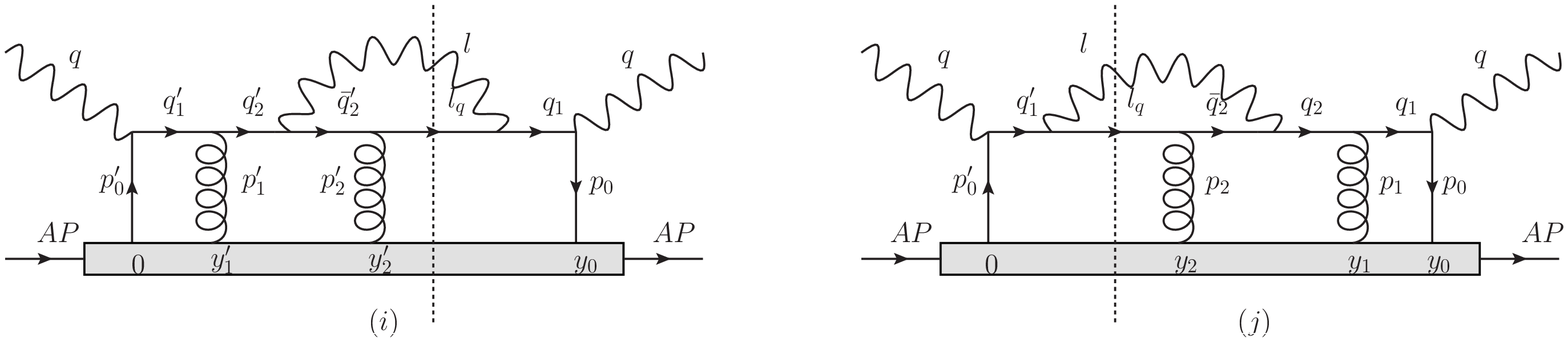}
\caption{Non-central-cut diagrams: two gluon insertions in the amplitude and zero in the complex conjugate (or vice versa). 
}
\label{20twist}
\end{figure}

To simplify the above expression, one may first isolate the phase factor associated with the photon insertion locations $z$ and $z'$: $e^{-i(\bar{q}_1 + l - q_1) \cdot z}  e^{i(\bar{q}_1' + l - q_1') \cdot z'}$.
After integrating out the locations $z$ and $z'$, one may obtain two $\delta$ functions to perform the integrations for the momenta $\bar{q}_1$ and $\bar{q}_1'$, yielding $\bar{q}_1= q_1-l$ and $\bar{q}_1'= q_1'- l$.
Using the momentum conservation at each interaction vertex, one can write down the following relations for various momenta in Fig. \ref{11twist}(a):
\begin{eqnarray}
p_0=q_1-q, \,\, p_1=l_q-\bar{q}_1; \,\,\,\,
p'_0=q'_1-q, \,\, p'_1=l_q-\bar{q}'_1.
\end{eqnarray}
From the above relations, one may change the integration variables: $q_1 \to p_0$ and $q'_1 \to p'_0$.
Re-introducing the four-momentum $p_1 = q + p_0 - l - l_q$, the phase factor for the hadronic tensor becomes: $e^{-ip_0\cdot y_0} e^{-ip_1\cdot y_1} e^{ip'_1\cdot y'_1}$.
In this work, we perform the calculation in the light-cone gauge ($A^-=0$) in the limit of very high energy and collinear radiation, therfore, the dominant component of the vector potential is the forward $(+)$-component, which means that only $(-)$-component of $\gamma$ matrices contribute.
In such limit, one may ignore the $(\perp)$-component of the quark field operators and factor out one-nucleon state from the nucleus state as follows:
\begin{eqnarray}
\langle A | \bar{\psi}(y_0) \gamma^\mu \hat{O} \gamma^\nu \psi(0) | A \rangle
\!\!
\approx \!\! A C_p^A \langle p | \bar{\psi}(y_0) \frac{\gamma^+}{2} \psi(0) | p \rangle
{\rm Tr} [\frac{\gamma^-}{2} \gamma^\mu \frac{\gamma^+}{2} \gamma^\nu]  {\rm Tr} [\frac{\gamma^-}{2} \langle A| \hat{O} |A\rangle].
\end{eqnarray}
With the above simplifications, the hadronic tensor can be written as the following form:
\begin{eqnarray}
W^{A\mu\nu}_{(a)} \!\!&=&\!\! \sum_q Q_q^4 e^2 g^2 \frac{1}{N_c} {\rm Tr}\left[T^{a_1} T^{a'_1}\right]
\int \frac{d^4l}{(2\pi)^4} (2\pi)\delta(l^2) \int \frac{d^4l_q}{(2\pi)^4} (2\pi)\delta(l_q^2)
\int d^4y_0 \int d^4y_1  \int d^4y'_1e^{-ip_0\cdot y_0} e^{-ip_1\cdot y_1} e^{ip'_1\cdot y'_1}
\nonumber\\ \!\!&\times&\!\!
\int \frac{d^4 p_0}{(2\pi)^4}  \int \frac{d^4 p'_0}{(2\pi)^4}
\int \frac{d^4p_1}{(2\pi)^4} (2\pi)^4 \delta^4(l+l_q - p_0 - p_1 - q)
\frac{1}{q_1^2-i\epsilon}  \frac{1}{\bar{q}_1^2-i\epsilon}
\frac{1}{q_1'^2+i\epsilon}  \frac{1}{\bar{q}_1'^2+i\epsilon}
\nonumber\\ \!\!&\times&\!\!(-g_\perp^{\mu\nu}) A C_p^A \langle p | \bar{\psi}(y_0) \frac{\gamma^+}{2} \psi(0) |p\rangle \langle A | A^{+a_1}(y_1) A^{+a'_1}(y'_1) |A\rangle
\times {\rm Tr}\left[\frac{\gamma^-}{2}  \slashed{q}_1 \gamma^\alpha \slashed{\bar{q}}_1 \gamma^- \slashed{l}_q   \gamma^- {\slashed{\bar{q}}_1'}
\gamma^\beta  \slashed{q}_1'\right] 
G_{\alpha\beta}(l) .
\end{eqnarray}
To further simplify the above expression, we look at the quark lines before and after the photon emission point,
\begin{eqnarray}
&& q_1^2 = (q+p_0)^2 = 2p^+q^-(1+x_0^-)[-x_B + x_0 - x_{D0}],
\nonumber\\
&& \bar{q}_1^2 = (q+p_0-l)^2 = 2p^+q^-(1+x_0^--y)[-x_B + x_0 - x_{C0}],
\end{eqnarray}
where we have defined the momentum fractions,
\begin{eqnarray}
&& \bar{x}_i = \sum_{j=0}^i x_j = \sum_{j=0}^i \frac{p_j^+}{p^+} ; \,
\bar{x}^-_i = \sum_{j=0}^i x^-_j = \sum_{j=0}^i \frac{p_j^-}{q^-}; \,
\nonumber\\
&&
\bar{x}_{Di} = \sum_{j=0}^i {x}_{Dj} = \frac{(\sum_{j=0}^i \mathbf{p}_{j\perp})^2}{2p^+q^-(1+ \bar{x}^-_i)}; \,
\bar{x}_{Ci} = \sum_{j=0}^i {x}_{Cj} = x_L(1-y) + \frac{( \sum_{j=0}^i \mathbf{p}_{j\perp} - \mathbf{l}_\perp)^2}{2p^+q^-(1+\bar{x}_i^--y)}.
\end{eqnarray}
The momenta $q_1'$ and $\bar{q}_1'$ may be treated in a similar way.
In terms of the above momentum fractions, the on-shell condition for the final outgoing quark $l_q$ gives,
\begin{eqnarray}
(2\pi)\delta(l_q^2)=\frac{1}{2p^+q^-(1+ \bar{x}_1^- - y)}(2\pi)\delta(-x_B + \bar{x}_1 - \bar{x}_{C1}).
\end{eqnarray}
Combining the contributions from the denominators from all the internal quark lines and the on-shell condition for the outgoing quark, one obtains:
\begin{eqnarray}
D_q \!\!&=&\!\! \frac{C_q}{(2p^+q^-)^5} (2\pi) \delta(-x_B + \bar{x}_1 - \bar{x}_{C1})
\nonumber\\\!\!&\times&\!\!
\frac{1}{-x_B+x_0-x_{D0}-i\epsilon}  \frac{1}{-x_B+x_0-x_{C0}-i\epsilon}
\frac{1}{-x_B+x_0'-x_{D0}' +i\epsilon}  \frac{1}{-x_B+ x_0'- x_{C0}'+i\epsilon}
\end{eqnarray}
where $C_q$ stands for
\begin{eqnarray}
C_q = \frac{1}{1+x_0^-} \frac{1}{1+x_0^--y} \frac{1}{1+ x_0'^-}\frac{1}{1+ x_0'^- -y} \frac{1}{1+ \bar{x}_1^- - y}
\end{eqnarray}
As for the numerators of the quark lines (the trace part), we note the following decomposition of the photon polarization sum $G_{\alpha \beta}$:
\begin{eqnarray}
 G^{+-} = G^{-+} = 0; \  G^{++} = \frac{2l^+}{l^-}; \  G^{\perp+} = G^{+\perp} = \frac{l_\perp}{l^-}; \ G_{\perp\perp}^{\alpha\beta} = -g_\perp^{\alpha\beta}.
\end{eqnarray}
One may perform contraction of the trace with various components of $G_{\alpha\beta}(l)$.
The terms combining with the projection $G_{\perp\perp}$ render:
\begin{eqnarray}
N_{\perp\perp} \!\!&=&\!\! \frac{2{(2q^-)}^3}{C_q} \left[\frac{\mathbf{p}_{0 \perp} \cdot \mathbf{p}_{0\perp}'}{(1+x_0^-)(1+x_0'^-)} + \frac{(\mathbf{p}_{0 \perp} -\mathbf{l}_\perp) \cdot (\mathbf{p}_{0\perp}'-\mathbf{l}_\perp)}{(1+x_0^--y)(1+x_0'^--y)}\right].
\end{eqnarray}
Similarly, the terms related to the projection $G^{++}$ give,
\begin{eqnarray}
N_{++} = \frac{4(2q^-)^3}{C_q} \frac{l_\perp^2}{y^2}.
\end{eqnarray}
The terms associated with the projections $G^{+\perp}$ and $G^{\perp+}$ read,
\begin{eqnarray}
N_{\perp+} + N_{+\perp} = -\frac{2(2q^-)^3}{C_q} \frac{1}{y}
\left[\frac{\mathbf{p}_{0 \perp} \cdot \mathbf{l}_{\perp}}{(1+x_0^-)} + \frac{\mathbf{p}_{0 \perp}' \cdot \mathbf{l}_{\perp}}{(1+x_0'^-)} + \frac{(\mathbf{p}_{0 \perp} -\mathbf{l}_\perp) \cdot \mathbf{l}_{\perp}}{(1+x_0^--y)} + \frac{(\mathbf{p}_{0 \perp}' -\mathbf{l}_\perp) \cdot \mathbf{l}_{\perp}}{ (1+x_0'^--y)}\right] .
\end{eqnarray}
After the above simplifications, the hadronic tensor takes the following form,
\begin{eqnarray}
W^{A\mu\nu}_{(a)} \!\!&=&\!\! \sum_q Q_q^4 e^2 g^2 \frac{1}{N_c} {\rm Tr}\left[ T^{a_1} T^{a_1'} \right]
\int \frac{d^4l}{(2\pi)^4} (2\pi)\delta(l^2) \int \frac{d^4l_q}{(2\pi)^4}   (2\pi)^4 \delta^4(l+l_q - p_0 - p_1 - q) \nonumber\\\!\!&\times&\!\!
\int d^4y_0 \int d^4y_1 \int d^4y_1'
\int \frac{d^3\mathbf{p}_0 dx_0}{(2\pi)^4}\int \frac{d^3\mathbf{p}_1 dx_1}{(2\pi)^4}\int \frac{d^3\mathbf{p}_0' dx_0'}{(2\pi)^4}
\nonumber\\\!\!&\times&\!\!
\left(e^{-i\mathbf{p}_0\cdot \mathbf{y}_0}  e^{-i\mathbf{p}_1\cdot \mathbf{y}_1}  e^{i\mathbf{p}_1'\cdot \mathbf{y}_1'}\right)
\left(e^{-ix_0 p^+y_0^-} e^{-ix_1 p^+y_1^-} e^{ix_1' p^+ y_1'^-} \right) (2\pi)\delta(-x_B + \bar{x}_1 - \bar{x}_{C1})
\nonumber\\\!\!&\times&\!\!
\frac{1}{-x_B+x_0-x_{D0}-i\epsilon}  \frac{1}{-x_B+x_0-x_{C0}-i\epsilon} \frac{1}{-x_B+x_0'-x_{D0}'+i\epsilon}  \frac{1}{-x_B+x_0'-x_{C0}'+i\epsilon}
\nonumber\\\!\!&\times&\!\!
(-g_\perp^{\mu\nu}) A C_p^A \langle p | \bar{\psi}(y_0) \frac{\gamma^+}{2} \psi(0) |p\rangle
 \langle A | A^{+a_1}(y_1) A^{+a'_1}(y'_1) |A\rangle  \nonumber \\ &&
\nonumber\\ \!\!&\times&\!\!
\frac{2}{(2p^+q^-)^2}
\frac{1 + \left(1 - \frac{y}{1+{x^-_0}} \right) \left(1 -  \frac{y}{1+{x_0'^-}} \right)}{y^2\left(1 - \frac{y}{1+{x^-_0}} \right) \left(1 -  \frac{y}{1+{x'^-_0}} \right)}
\left(\mathbf{l}_\perp - \frac{y}{1+{x^-_0}} \mathbf{p}_{0\perp} \right)\cdot \left(\mathbf{l}_\perp -\frac{y}{1+{x'^-_0}} \mathbf{p}_{0\perp}' \right).
\end{eqnarray}
In the above expression, we have changed the integration variables $p_i^+ \to x_i$ and $p_j'^+ \to x'_j$.
For easy writing, the three-vector notations are utilized here for the momentum and coordinate space variables: $\mathbf{p} = (p^-, \mathbf{p}_\perp)$ and $\mathbf{y} = (y^+, \mathbf{y}_\perp)$; their dot product reads as: $\mathbf{p}\cdot \mathbf{y} = p^-y^+ - \mathbf{p}_\perp \cdot \mathbf{y}_\perp$.

Now we perform the integration over the momentum fractions $x_0$, $x_1$ and $x_0'$. The corresponding phase factors read as follows:
\begin{eqnarray}
\Gamma^+ =e^{-ix_0 p^+y_0^-} e^{-ix_1 p^+y_1^-} e^{i x'_1p^+ y_1'^-}=e^{-ix_0p^+(y_0^--{y'_1}^-)} e^{-ix_1p^+(y_1^--{y'_1}^-)} e^{ix_0' p^+(y_0'^-- y_1'^-)}
\end{eqnarray}
where we have used the overall momentum conservation $p'_1 = p_0 + p_1 -  p'_0$.
One may first integrate out the momentum fraction $x_1$ using the on-shell condition $2\pi \delta(-x_B + \bar{x}_1 - \bar{x}_{C1})$ for the outgoing quark line, then the phase factor becomes:
\begin{eqnarray}
\Gamma^+ = \left(e^{-i(x_B + \bar{x}_{C1})p^+y_1^-}e^{-ix_0p^+(y_0^--y_1^-)}\right) \left(e^{i(x_B + \bar{x}_{C1}')p^+y_1'^-}e^{ix'_0p^+(y_0'^--y_1'^-)} \right) = \Gamma_1^+ \Gamma_1'^+, 
\end{eqnarray}
where $\Gamma_1^+$ and $\Gamma_1'^+$ denote the the parts associated with the amplitude and the complex conjugate.
With the help of the contour integration technique, the integrations over the momentum fractions $x_0$ and $x'_0$ may be performed.
The $x_0$ integral may be carried out by closing the contour with a counter-clockwise semi-circle in the upper half of the $x_0$ complex plane:
\begin{eqnarray}
&&
\int \frac{dx_0}{2\pi}  \frac{e^{-ix_0 p^+(y_0^--y_1^-)}}{(-x_B + x_0 - x_{C0} - i \epsilon)(-x_B + x_0 - x_{D0} - i \epsilon)}
\nonumber\\
&&
= {i\theta(y_1^--y_0^-)   e^{-ix_B p^+(y_0^--y_1^-)}}
\frac{ e^{-i x_{C0} p^+ (y_0^- - y_1^-)} - e^{-i x_{D0} p^+ (y_0^- - y_1^-)}} {x_{C0} - x_{D0}}.
\end{eqnarray}
Two terms in the equation represent the fact that each one of the propagators is close to on-shell, and the $\theta$-function means that the quark line propagates from $y_0^-$ to $y_1^-$.
We can now collect the phase factor as follows:
\begin{eqnarray}
\Gamma_1^+ \to  i\theta(y_1^- - y_0^-) e^{-i x_B p^+ y_0^-}e^{-ix_{D0}p^+y_0^-}e^{-i x_{C1} p^+ y_1^-}
\frac{e^{-i\delta x_{D0} p^+ y_0^-} - e^{-i\delta x_{D0} p^+ y_1^-}} {\delta x_{D0}}, \ \
\end{eqnarray}
where for convenience we have defined the momentum fraction variables $\delta x_{Di} = x_{Ci} - x_{Di}$ (as well as $\delta \bar{x}_{Di} = \bar{x}_{Ci} - \bar{x}_{Di}$,  $\delta x_{Di}' = x_{Ci}' - x_{Di}'$,  and $\delta \bar{x}_{Di}' = \bar{x}_{Ci}' - \bar{x}_{Di}'$).
The integration over $x'_0$ is completely analogous, except that we close the contour with a clockwise semi-circle in the upper half of the complex plane, and associated with the $\theta$ function is a factor of $(-i)$ instead of $i$.
After the integrations over the momentum fractions $x_0$, $x_1$ and $x_0'$ in the quark lines have been done, the hadronic tensor reads as follows:
\begin{eqnarray}
W^{A\mu\nu}_{(a)}  \!\!&=&\!\! \sum_q Q_q^4 e^2 g^2 \frac{1}{N_c} {\rm Tr}\left[ T^{a_1} T^{a'_1} \right]
\int \frac{d^4l}{(2\pi)^4} (2\pi)\delta(l^2) \int \frac{d^4l_q}{(2\pi)^4}   (2\pi)^4 \delta^4(l+l_q - p_0 - p_1 - q)
\\\!\!&\times&\!\!
\int d^4y_0 \int d^4y_1 \int d^4y'_1
\int \frac{d^3\mathbf{p}_0}{(2\pi)^3}\int \frac{d^3\mathbf{p}_1 }{(2\pi)^3}\int \frac{d^3\mathbf{p}'_1}{(2\pi)^3}
\left(e^{-i\mathbf{p}_0\cdot \mathbf{y}_0}  e^{-i\mathbf{p}_1\cdot \mathbf{y}_1}  e^{i\mathbf{p}_1'\cdot \mathbf{y}_1'} \right) \nonumber\\\!\!&\times&\!\!
 (-g_\perp^{\mu\nu}) A C_p^A \langle p | \bar{\psi}(y_0) \frac{\gamma^+}{2} \psi(0) |p\rangle  \langle A |  A^{+a_1}(y_1)A^{+a'_1}(y'_1) |A\rangle
  \left( \theta(y_1^- - y_0^-)\theta(y_1'^- - y_0'^-)\right) e^{-ix_Bp^+y_0^-}
 \nonumber \\ \!\!&\times&\!\!
 \left(e^{-ix_{D0}p^+y_0^-} e^{-i x_{C1} p^+ y_1^-}e^{ix_{C1}' p^+ y_1'^-}\right)\left(e^{-i\delta x_{D0} p^+ y_0^-} - e^{-i\delta x_{D0} p^+ y_1^-}\right)
\left( 1 - e^{i\delta x_{D0}' p^+ y_1'^-}\right)
\nonumber\\ \!\!&\times&\!\!
\frac{1}{\delta x_{D0} \delta x_{D0}'}
\frac{2}{(2p^+q^-)^2}
\frac{1 + \left(1 - \frac{y}{1+ x^-_0} \right) \left(1 -  \frac{y}{1+ x_0'^-} \right)}{y^2\left(1 - \frac{y}{1+ x^-_0} \right) \left(1 -  \frac{y}{1+ x_0'^-} \right)}
\left(\mathbf{l}_\perp - \frac{y}{1+ x^-_0} \mathbf{p}_{0\perp} \right)\cdot \left(\mathbf{l}_\perp -\frac{y}{1+x_0'^-} \mathbf{p}_{0\perp}' \right), \nonumber
\end{eqnarray}
where we have changed the integral variable: $p_0' \to p_1'$ and set the location $y_0' = 0$.
Assuming that in the limit of very high energy, the nucleus may be approximated by a weakly interacting homogenous gas of nucleons, we can simplify the matrix elements of the gluon vector potentials in the nucleus state (note that the quark operators has already been factorized out).  
Since a nucleon is a color singlet, one may express the expectation of field operators in the nucleus states in terms of the expectations in the nucleon states as follows:
\begin{eqnarray}
 \langle A |  A^{+a_1}(y_1) A^{+a_1'}(y_1')  |A\rangle
= \left(\frac{\rho}{2p^+}\right)  \frac{\delta_{a_1 a_1'}}{N_c^2 - 1} \langle p| A^+(y_1) A^+(y_1') | p\rangle,
\end{eqnarray}
where $\rho$ is the nucleon density inside the nucleus.
Also in the above expression, we have averaged the colors of gluon field operators.
With the above simplification, the trace for the color matrices can be easily evaluated,
\begin{eqnarray}
\frac{1}{N_c} \delta_{a_1 a'_1} {\rm Tr}\left[ T^{a_1} T^{a'_1} \right] = C_F.
\end{eqnarray}
For conveinence, we change the four-vector location variables $(y_1, y_1')\rightarrow (Y_1, \delta y_1)$,
\begin{eqnarray}
Y_1 = \frac{1}{2}(y_1 + y_1');  &&  \delta y_1 = y_1 - y_1'.
\end{eqnarray}
Using the translational invariance of the correlation functions,
$\langle p| A^+(y_1) A^+(y'_1) | p\rangle \approx \langle p| A^+(\delta{y_1}) A^+(0) | p\rangle$,
the integration for the phase factor can be performed:
\begin{eqnarray}
\int d^3\mathbf{y}_1 \int d^3\mathbf{y}_1' e^{-i\mathbf{p}_1 \cdot \mathbf{y}_1} e^{i\mathbf{p}_1'\cdot \mathbf{y}_1'}
= (2\pi)^3 \delta^3(\mathbf{p}_1 - \mathbf{p}_1') \int d^3\delta \mathbf{y}_1 e^{-i(\mathbf{p}_1+\mathbf{p}_1')\cdot \frac{\delta\mathbf{y}_1}{2}}.
\end{eqnarray}
In the above expression, the $\delta$ function represents the fact that the two gluon field insertions in the amplitude and the complex cojugate carry the same momentum, $\mathbf{p}_1' = \mathbf{p}_1$, which also implies $\mathbf{p}_0' = \mathbf{p}_0$.
The $\delta$ functions may be utilized to carry out the integration over the momentum $\mathbf{p}'_1$.
Recalling the expressions for the momentum fractions $x_{D0}$ and $x_{C0}$, one may obtain the momentum fractions $\delta x_{D0}$ as follows:
\begin{eqnarray}
\delta x_{D0} = x_{C0} - x_{D0}
= \frac{\left(\mathbf{l}_\perp - \frac{y }{1+{x_0^-}} \mathbf{p}_{0\perp} \right)^2}{2p^+q^-y(1-\frac{y}{1+{x_0^-}})}= x_L.
\end{eqnarray}
The expressions for $\delta x_{D0}'$ are completely analogous. 
We may further perform the integrations over the momentum $\mathbf{p}_0$ and the location $\mathbf{y}_0$, rendering $\mathbf{p}_0=0$.
Using the above results, one may obtain the hard part of the matrix element (the last line of the hadronic tensor, denoted as $T_{(a)}$) as:
\begin{eqnarray}
\label{linear_approximation}
T_{(a)} \!\!&=&\!\! \frac{2yP(y)}{l_\perp^2}.
\end{eqnarray}
Now we simplify the phase factor (the second last line in the hadronic tensor).
Keeping only the leading contribution, 
the phase factor (denoted as  $S_{(a)}$) can be obtained as:
\begin{eqnarray}
S_{(a)} \approx e^{-ix_Lp^+y_0^-}(1- e^{-i x_L p^+ (Y_1^-+\frac{1}{2}\delta y_1^--y_0^-)})( 1 - e^{ix_L p^+ (Y_1^--\frac{1}{2}\delta y_1^-)}).
\end{eqnarray}
We further note that $Y_1^-$ is the location of the photon insertion location which spans over the size of the nucleus, while $y_0^-$ and $\delta y_1^-$ are confined within the size of one nucleon.
Thus, $y_0^-$ and $\delta y_1^-$ is much smaller than $Y_1^-$ and the phase factor may be further approximated as:
\begin{eqnarray}
S_{(a)} \approx e^{-ix_Lp^+y_0^-} [2-2 \cos(x_L p^+ Y_1^-)].
\end{eqnarray}
With the above simplifications, the hadronic tensor reads as follows:
\begin{eqnarray}
W^{A\mu\nu}_{(a)} \!\!&=&\!\! (-g_\perp^{\mu\nu}) A C_p^A \sum_q Q_q^4 \frac{\alpha_e}{2\pi}
\int dy P(y) \int \frac{d^2l_\perp}{\pi l_\perp^2} \int{d^3\mathbf{l}_q}
\int dy_0^- e^{-i(x_B+x_L)p^+y_0^-}  \langle p | \bar{\psi}(y_0^-) \frac{\gamma^+}{2} \psi(0) |p\rangle
\nonumber\\ \!\!&\times&\!\!
\int d Y_1^- \int d \delta y_1^- \left(g^2 \frac{C_F}{N_c^2-1}\frac{\rho}{2 p^+}\right) \int d^3\delta \mathbf{y}_1 \int \frac{d^3\mathbf{p}_1}{(2\pi)^3} e^{-i\mathbf{p}_1\cdot \delta \mathbf{y}_1} \langle p| A^+(\delta{y_1}) A^+(0) | p\rangle
\nonumber\\ \!\!&\times&\!\!
[2-2 \cos(x_L p^+ Y_1^-)] \delta^3(\mathbf{l}+\mathbf{l}_q - \mathbf{p}_1 - \mathbf{q}).
\end{eqnarray}
The above expression is the contribution to the hadronic tensor from Fig.\ref{11twist}(a). The calculations for other diagrams in Fig. \ref{11twist} and Fig. \ref{20twist} are completely analogous; their main results are provided in the Appendix.
Summing over the contributions from all central-cut diagrams in Fig. \ref{11twist} and carrying out the integral $\int_0^{L^-} dY_1^-$ (assuming a homogeneous medium and the transport coefficients $D_{L1}$, $D_{L2}$ and $D_{T2}$ are position independent), the hadronic tensor contributed from all central-cut diagrams reads as follows:
\begin{eqnarray}
\label{W11_munu}
W^{A\mu\nu}_{(abcd)} &=& (-g_\perp^{\mu\nu}) A C_p^A \sum_q Q_q^4 \frac{\alpha_e}{2\pi}
\int dy P(y) \int \frac{d^2l_\perp}{\pi l_\perp^2} \int{d^3\mathbf{l}_q} (2\pi) f_q(x_B+x_L) \nonumber\\\!\!&\times&\!\!
L^- \int d\delta y_1^- \left(g^2\frac{C_F}{N_c^2 - 1}\frac{\rho}{2p^+}\right) \int d^3\delta \mathbf{y}_1 \int \frac{d^3\mathbf{p}_1}{(2\pi)^3} e^{-i\mathbf{p}_1\cdot \delta \mathbf{y}_1} \langle p | A^+(\delta y_1^-, \delta \mathbf{y}_1) A^+(0)  |p\rangle
\nonumber \\ &\times&
\left[1 + \frac{y(1-y)}{1+(1-y)^2} F_L \frac{{p_1}^-}{q^-} + y F_L \frac{\mathbf{l}_\perp \cdot \mathbf{p}_{1\perp}}{l^2_\perp} + \frac{y}{1+(1-y)^2} \left\{y- (1-y)F_L\right\} \left(\frac{p_1^-}{q^-}\right)^2 + y^2 (1 - F_L)\frac{p_{1\perp}^2}{l^2_\perp}
\nonumber \right. \\ &&
\left.
+ 2y^2 F_L \frac{(\mathbf{l}_\perp \cdot \mathbf{p}_{1\perp})^2}{l^4_\perp}
+ \left\{\frac{y^2(1-y)}{1+(1-y)^2}(2+F_L) - y F_L\right\}\frac{{p_1}^-}{q^-} \cdot \frac{\mathbf{l}_\perp \cdot \mathbf{p}_{1\perp}}{l^2_\perp}\right]
\delta^3(\mathbf{l}+\mathbf{l}_q - \mathbf{p}_1 - \mathbf{q}).
\end{eqnarray}
where
\begin{eqnarray}
F_L = \frac{2\sin(x_L p^+ L^-)}{x_L p^+ L^-}
\end{eqnarray}
In the above expression, we have kept terms in the hard parts of matrix elements up to the second order in $\frac{p_1^-}{q^-}$ and $\frac{{p}_{1\perp}}{{l}_\perp}$ and their cross terms, to be consistent with the momentum gradient expression up to the second order as will be done in a short moment.
Summing over the contributions from all non-central cut diagrams in Fig. \ref{20twist} and carrying out the integral $\int_0^{L^-} dY_1^-$, the hadronic tensor contributed from all non-centra-cut diagrams read as follows:
\begin{eqnarray}
W^{A\mu\nu}_{(efdghi)} \!\!&\approx&\!\! (-g_\perp^{\mu\nu}) A C_p^A\sum_q Q_q^4 \frac{\alpha_e}{2\pi} \int dy P(y) \int \frac{d^2l_{\perp}}{\pi l^2_{\perp}}
\int d^3\mathbf{l}_q  (2\pi) f_q(x_B+x_L)
\nonumber\\\!\!&\times&\!\!
(-L^-) \int d \delta y_1^-  \left(g^2 \frac{C_F}{N^2_c-1} \frac{\rho}{2 p^+}\right) \langle p | A^+(\delta y_1^-) A^+(0)  |p\rangle
\nonumber\\\!\!&\times&\!\!
\delta^3(\mathbf{l}+\mathbf{l}_q - \mathbf{q}).
\end{eqnarray}
To do further simplification, we introduce the momentum gradient expansion for the hard part $H(\mathbf{p}_1)$ (the last two lines of $W^{A\mu\nu}_{(abcd)}$ and the last line of $W^{A\mu\nu}_{(efdghi)}$).
Assuming the momentum exchange between the hard parton and medium constituents is small, one may expand $H(\mathbf{p}_1)$ as a series of the Taylor expansion in the three-dimensional momenta $\mathbf{p}_1 = (p_1^-, \mathbf{p}_{1\perp})$ around $\mathbf{p}_1 \to 0$:
\begin{eqnarray}
\label{gradient_expansion_1}
H = \left[1 + p_1^\alpha \frac{\partial}{\partial p_1^\alpha} + \frac{1}{2} p_1^\alpha p_1^\beta \frac{\partial}{\partial p_1^\alpha} \frac{\partial}{\partial p_1^\beta} + \cdots \right] H|_{\mathbf{p}_1 = 0}, \ \
\end{eqnarray}
where $\alpha$ and $\beta$ take the values: ``$-$'' or ``$\perp$".
In the above expansion, we have only kept the terms up to the second order derivative in both longitudinal and transverse momenta.
High-order terms are neglected and should be straightforward to include in the gradient expansion; they correspond to higher-order moments for the momentum distribution of the exchanged gluons.
The zeroth order term (without momentum derivative) represents the case where the exchanged gluon carry zero momenta; it contributes to the gauge corrections to the leading twist results, and will not be considered further \cite{Luo:1994np, Wang:2001ifa, Zhang:2003bsa, Zhang:2004qm}.
One may notice that non-central-cut diagrams only contribute to the zeroth order term in the above gradient expansion, thus do not physically contribute to medium-induced single photon radiation.
Now performing the integration over $\mathbf{p}_1$ by part, one may obtain:
\begin{eqnarray}
&&\langle p| A^+(\delta \mathbf{y}_1) A^+(0)|p \rangle e^{-\mathbf{p}_1 \cdot \delta\mathbf{y}_1} p_1^\alpha \frac{\partial}{\partial p_1^\alpha} =   e^{-\mathbf{p}_1 \cdot \delta\mathbf{y}_1} (-i) \langle p| \partial^\alpha A^+(\delta \mathbf{y}_1) A^+(0)|p \rangle \frac{\partial}{\partial p_1^\alpha},  \\
&&\langle p| A^+(\delta \mathbf{y}_1) A^+(0)|p \rangle e^{-\mathbf{p}_1 \cdot \delta\mathbf{y}_1} p_1^\alpha p_1^\beta \frac{\partial}{\partial p_1^\alpha} \frac{\partial}{\partial p_1^\beta} = e^{-\mathbf{p}_1 \cdot \delta\mathbf{y}_1} \langle p| \partial^\alpha A^+(\delta \mathbf{y}_1) \partial^\beta A^+(0)|p \rangle \frac{\partial}{\partial p_1^\alpha} \frac{\partial}{\partial p_1^\beta}.
\end{eqnarray}
We note the condition $\mathbf{p}_1 \to 0$ imposed by the Taylor expansion.
This means that the hard part $H(\mathbf{p}_1)$ no longer depends on $\mathbf{p}_1$, thus the integrations for $\mathbf{p}_1$ and $\delta\mathbf{y}_1$ may be directly carried out.
Then the hadronic tensor may be written as follows:
\begin{eqnarray}
W^{A\mu\nu} \!\!&=&\!\! (-g_\perp^{\mu\nu}) A C_p^A \sum_q Q_q^4 \frac{\alpha_e}{2\pi}
\int dy P(y) \int \frac{d^2l_\perp}{\pi l_\perp^2}  \int{d^3\mathbf{l}_q} (2\pi) f_q(x_B+x_L) \phi(L^-, l_q^-, \mathbf{l}_{q\perp})
\end{eqnarray}
where
\begin{eqnarray}
\phi(L^-, l_q^-, \mathbf{l}_{q\perp}) \!\!&=&\!\!
L^-  \left[ - D_{L1}\frac{\partial}{\partial p_1^-} + \frac{1}{2} D_{L2} \frac{\partial^2}{\partial^2 p_1^-} + \frac{1}{2} D_{T2} {\nabla p_{1 \perp}^2}  \right]
 H|_{\mathbf{p}_1 = 0}.
\end{eqnarray}
Here we consider the unpolarized initial and final states and only keep the terms with non-vanishing coefficients by using the symmetry of the system.
The parton transport coefficients $D_{L1}$, $D_{L2}$, $D_{T2}$ are defined as follows:
\begin{eqnarray}
D_{L1} \!\!&=&\!\! g^2 \frac{C_F}{N_c^2 - 1} \int dy^- \frac{\rho}{2p^+} \langle p| i\partial^- A^+(y^-) A^+(0)|p \rangle, \\
D_{L2} \!\!&=&\!\! g^2 \frac{C_F}{N_c^2 - 1} \int dy^- \frac{\rho}{2p^+} \langle p| \partial^- A^+(y^-) \partial^- A^+(0)|p \rangle, \\
D_{T2} \!\!&=&\!\! g^2 \frac{C_F}{N_c^2 - 1} \int dy^- \frac{\rho}{2p^+} \langle p| \partial_\perp A^+(y^-) \partial_\perp A^+(0)|p \rangle. \ \
\end{eqnarray}
Up to an overall factor, the transport coefficients $D_{L1}$, $D_{L2}$ and $D_{T2}$ are equivalent to $\hat{e}$ (the longitudinal momentum loss rate), $\hat{e}_2$ (longitudinal momentum diffusionrate), and $\hat{q}$ (the transverse momenta broadening rate), respectively \cite{Qin:2012fua}.
We should point out that the above definitions of parton transport coefficients are not gauge-invariant.
To obtain the gauge-invariant definitions of these transport coefficients, one needs to resum all the diagrams with arbitrary numbers of soft gluon insertions, which gives the Wilson lines between the gluon field operators.
One may find more discussions on gauge-invariant definition of $\hat{q}$ in Ref. \cite{Liang:2008rf, Benzke:2012sz}.

We now analyze the seven terms in the hard part $H$ one by one:
\begin{eqnarray}
H = H_0 + H_-^{(1)} + H_\perp^{(1)} + H_-^{(2)} + H_{\perp, 1}^{(2)} + H_{\perp, 2}^{(2)} +H_{-\perp}^{(2)}.
\end{eqnarray}
Accordingly, the distribution function $\phi(L^-, l_q^-, \mathbf{l}_{q\perp})$ is also splitted into seven contributions:
\begin{eqnarray}
\phi = \phi^{(0)} + \phi_-^{(1)} + \phi_\perp^{(1)} + \phi_-^{(2)} + \phi_{\perp, 1}^{(2)} + \phi_{\perp, 2}^{(2)} + \phi_{-\perp}^{(2)}.
\end{eqnarray}
One can directly read off the contribution from the term $H_0$:
\begin{eqnarray}
\phi^{(0)}
\!\!&=&\!\!L^- \left[ D_{L1}\frac{\partial}{\partial l_q^-} + \frac{1}{2} D_{L2} \frac{\partial^2}{\partial^2 l_q^-} + \frac{1}{2} D_{T2} {\nabla_{l_{q\perp}}^2}  \right] \delta(l_q^- - q^-(1-y)) \delta^2(\mathbf{l}_{q\perp} + \mathbf{l}_{\perp}),
\end{eqnarray}
where the derivative over $p_1$ has been converted to the derivative over $l_q$.
The distribution function $\phi_-^{(1)}(L^-, l_q^-, \mathbf{l}_{q\perp})$ reads as:
\begin{eqnarray}
\phi_-^{(1)} \!\!&=&\!\! L^- \left[ - D_{L1}\frac{\partial}{\partial p_1^-} + \frac{1}{2} D_{L2} \frac{\partial^2}{\partial^2 p_1^-}  + \frac{1}{2} D_{T2} {\nabla p_{1 \perp}^2}  \right] \left\{\frac{y(1-y)}{1+(1-y)^2}  F_L  \frac{{p_1}^-}{q^-} \delta^3(\mathbf{l}+\mathbf{l}_q - \mathbf{p}_1 - \mathbf{q})\right\}|_{\mathbf{p}_1 = 0}
\end{eqnarray}
Performing the derivative over $p_1$ and then converting it to the derivative over $l_q$, one obtains:
\begin{eqnarray}
\phi_-^{(1)}
\!\!&=&\!\!\frac{y(1-y)}{1+(1-y)^2}  F_L  \left(-\frac{D_{L1}L^-}{q^-} - \frac{D_{L2}L^-}{q^-} \frac{\partial}{\partial l_q^-} \right) \delta(l_q^- - q^-(1-y)) \delta^2(\mathbf{l}_{q\perp} + \mathbf{l}_{\perp}).
\end{eqnarray}
The analysis for other five terms are completely analogous, and they read as:
\begin{eqnarray}
\phi_\perp^{(1)} &=& y  F_L   \left(-\frac{ D_{T2}L^-}{l_\perp^2}  \mathbf{l}_\perp \cdot \nabla_{l_{q\perp}}\right) \delta(l_q^- - q^-(1-y)) \delta^2(\mathbf{l}_{q\perp} + \mathbf{l}_{\perp});
\nonumber\\
\phi_-^{(2)} &=& \frac{y}{1+(1-y)^2} \left\{y - (1-y) F_L\right\} \frac{D_{L2} L^-}{{{q^-}^2}}\delta(l_q^- - q^-(1-y)) \delta^2(\mathbf{l}_{q\perp} + \mathbf{l}_{\perp});
\nonumber\\
\phi_{\perp,1}^{(2)} &=& 2y^2 (1 - F_L)  \frac{D_{T2}L^-}{{l_\perp^2}} \delta(l_q^- - q^-(1-y)) \delta^2(\mathbf{l}_{q\perp} + \mathbf{l}_{\perp});
\nonumber\\
\phi_{\perp,2}^{(2)} &=& 2y^2 F_L \frac{D_{T2}L^-}{ l_\perp^2}\delta(l_q^- - q^-(1-y)) \delta^2(\mathbf{l}_{q\perp} + \mathbf{l}_{\perp});
\nonumber\\
\phi_{-\perp}^{(2)} &=& 0.
\end{eqnarray}
Putting seven contributions together, the hadronic tensor takes the following form:
\begin{eqnarray}
W^{A \mu\nu} \!\!&=&\!\!
(-g_\perp^{\mu\nu}) A C_p^A \sum_q Q_q^4 \frac{\alpha_e}{2\pi}
\int dy P(y)\int \frac{d^2 l_\perp}{\pi l_\perp^2}\int d^3 \mathbf{l}_q (2\pi) f_q(x_B + x_L)
\nonumber\\ \!\!&\times&\!\!
 \left\{ \left[ D_{L1} L^-\frac{\partial}{\partial l_q^-} + \frac{1}{2} D_{L2} L^- \frac{\partial^2}{\partial^2 l_q^-} + \frac{1}{2} D_{T2} L^- {\nabla_{l_{q\perp}}^2}\right] + \frac{y(1-y)}{1+(1-y)^2} F_L  \left(-\frac{D_{L1} L^-}{q^-} - \frac{D_{L2} L^-}{q^-} \frac{\partial}{\partial l_q^-} \right)
\nonumber\right.\\\!\!&&\!\!
\left.  + y F_L \left(- D_{T2} L^- \frac{\mathbf{l}_\perp \cdot \nabla_{l_{q\perp}}}{l_\perp^2} \right) + \frac{y}{1+(1-y)^2} \left[y - (1-y) F_L \right] \frac{D_{L2} L^-}{{{q^-}^2}}
+ 2y^2\frac{D_{T2} L^-}{l_\perp^2} \right\}
\nonumber\\
\!\!&\times&\!\!
 \delta(l_q^- - q^-(1-y)) \delta^2(\mathbf{l}_{q\perp} + \mathbf{l}_{\perp})
\end{eqnarray}
In the extremely high energy and collinear limits where the exchanged momentum between the hard parton and medium is small, one can perform the integration over the three-dimensional momentum of the final outgoing quark $\mathbf{l}_q=(l_q^-,\mathbf{l}_{q\perp})$, and obtain the single-differential hadronic tensor as follows:
\begin{eqnarray}
\frac{dW^{A \mu\nu}}{dy d{l^2_{\perp}}} \!\!&=&\!\!  (-g_\perp^{\mu\nu}) A C_p^A \sum_q Q_q^4 \frac{\alpha_e}{2\pi}\frac{P(y)}{l_\perp^2} (2\pi) f_q(x_B + x_L)
\nonumber\\\!\!&\times&\!\!
\left\{ - \frac{y(1-y)}{1+(1-y)^2} F_L \frac{D_{L1}L^-}{q^-} + \frac{y}{1+(1-y)^2} \left[y - (1-y) F_L \right] \frac{D_{L2}L^-}{{{q^-}^2}} + 2y^2\frac{D_{T2} L^-}{l_\perp^2}\right\}
\end{eqnarray}
From the above expression, one may read off the medium-induced photon bremsstrahlung spectrum:
\begin{eqnarray}
\label{dNgamma_11}
\frac{dN^{\rm med}_{\gamma}}{dy d{l^2_{\perp}}} \!\!&=&\!\! \frac{\alpha_e}{2\pi} \frac{P(y)}{l_\perp^2}
\left\{ - \frac{y(1-y)}{1+(1-y)^2} F_L \frac{D_{L1}L^-}{q^-} + \frac{y}{1+(1-y)^2} \left[y - (1-y) F_L \right] \frac{D_{L2}L^-}{{{q^-}^2}} + 2y^2\frac{D_{T2} L^-}{l_\perp^2}\right\},
\end{eqnarray}
\end{widetext}
where $y$ is the fraction of the forward momentum carried by the radiated photon from the quark, and $l_\perp$ is the transverse momentum of the radiated photon.
One can clearly see the individual contributions from the drag and the diffusions of the longitudinal momentum and the transverse momentum diffusion to the single photon bremsstrahlung spectrum.
A notable result is that unlike the transverse momentum broadening which induces additional radiation in medium, the longitudinal momentum loss (the drag) tends to suppress the medium-induced photon radiation.
Another interesting observation is that the contributions from the drag and diffusions decouple for the single photon emission spectrum.
This is due to the fact that we have only kept the terms up to the second order in the momentum gradient expansion. 
In principle, one would expect to receive the contribution from the coupling between the drag and diffusions of the hard parton's momentum when higher order terms are included in the momentum gradient expansion.

\section{Photon bremsstrahlung from multiple scatterings}

In the previous section, we have computed medium-induced photon radiation spectrum from a hard quark undergoing a single scattering from the dense medium, which corresponds to the contribution involving twist-four parton distribution in the nucleus.
In this section, we consider the photon bremsstrahlung process from a quark jet which experiences multiple scatterings with the medium constituents.
This corresponds to higher twist contribution, i.e., containing more partonic operators in the medium.
Higher twist contribution is usually suppressed by powers of the hard scale $Q^2$, but a sub-class of these contributions may be length-enhanced for an extended medium \cite{Guo:2000nz, Wang:2001ifa, Majumder:2009ge}. In this section, we we will isolate and resum the length-enhanced higher twist contribution to the photon bremsstrahlung process from a hard quark jet.
We also compare the result from the resummation of multiple scatterings to that for single scattering.

Fig.~\ref{mntwist} shows the generic diagram for the photon bremsstrahlung process from a hard quark jet undergoing multiple scatterings in the dense nuclear medium ($m$ scatterings in the amplitude and $n$ scatterings in the complex conjugate).
A hard virtual photon strikes a quark with momentum $p_0'$ in the amplitude ($p_0$ in the complex conjugate) in the nucleus at the location $y_0'=0$ in the amplitude ($y_0$ in the complex conjugate), and the struck quark is then sent back to the nucleus.
During its propagation, the hard quark with momentum $q_1'$ ($q_1$ in the complex conjugate) scatters off the gluon fields in the nucleus at the locations $y_j'$ with $0<j<m$ ($y_i$ in the complex conjugate with $0<i<n$), and picks up momentum $p_j'$ ($p_i$ in the complex conjugate) via each scattering with the medium constituents.
The photon with momentum $l$ is radiated from the hard quark at the location $z$ between the locations $y_q$ and $y_{q+1}$ (at the location $z'$ between the locations $y_p$ and $y_{p+1}$ in the complex conjugate).
Here the momenta of the quark lines after the scattering at the location $y'_j$ are denoted as: $q'_{j+1}$ with $j \le q$ before the photon emission, and $\bar{q}'_{j+1}$ with $j>q$ after the photon emission (at the location $y_i$: $q_{i+1}$ with $i\le p$, and $\bar{q}_{i+1}$ with $i>p$, respectively, in the complex conjugate).
The quark lines before and after the photon radiation vertex are denoted as $q'_{q+1}$ and $\bar{q}'_{q+1} = q'_{q+1}-l$ ($q_{p+1}$ and $\bar{q}_{p+1} = q_{p+1}-l$ in the complex conjugate).
The final outgoing quark carries the momentum $l_q$.
One may write down the hadronic tensor for Fig. \ref{mntwist} as follows:
\begin{widetext}
\begin{eqnarray}
W^{A\mu\nu}_{nmpq} \!\!&=&\!\! \sum_q Q_q^4 e^2 g^{n+m} \frac{1}{N_c} {\rm Tr}\left[\left(\prod_{i=1}^{n} T^{a_i}\right) \left(\prod_{j=m}^{1} T^{a'_j}\right) \right]
\int \frac{d^4l}{(2\pi)^4} (2\pi)\delta(l^2) \int \frac{d^4l_q}{(2\pi)^4} (2\pi)\delta(l_q^2) \nonumber\\\!\!&\times&\!\!
\int d^4y_0 e^{iq\cdot y_0} \left(\prod_{i=1}^{n} \int d^4y_i\right) \left(\prod_{j=1}^{m} \int d^4y'_j \right) \int d^4z \int d^4z'
\nonumber \\\!\!&\times&\!\!
\left(\prod_{i=1}^{p} \int \frac{d^4q_i}{(2\pi)^4} e^{-iq_i\cdot (y_{i-1} - y_i)} \right) \left( \int \frac{d^4q_{p+1}}{(2\pi)^4} e^{-i q_{p+1} \cdot (y_p-z)} e^{-il\cdot z} \int \frac{d^4\bar{q}_{p+1}}{(2\pi)^4} e^{-i\bar{q}_{p+1}\cdot (z-y_{p+1})} \right) \nonumber\\\!\!&\times&\!\!
\left(\prod_{i=p+2}^{n} \int \frac{d^4\bar{q}_i}{(2\pi)^4} e^{-i\bar{q}_i\cdot (y_{i-1} - y_i)} \right) e^{-il_q \cdot (y_{n} - y_m')}
\left(\prod_{j=q+2}^{m}  \int \frac{d^4\bar{q}'_j}{(2\pi)^4} e^{-i\bar{q}_j' \cdot (y_j' - y_{j-1}')}\right)
\nonumber \\ \!\!&\times&\!\!
\left( \int \frac{d^4\bar{q}_{q+1}'}{(2\pi)^4} e^{-i \bar{q}_{q+1}' \cdot (y_{q+1}'-z')} e^{il\cdot z'} \int \frac{d^4q_{q+1}'}{(2\pi)^4} e^{-iq_{q+1}'\cdot (z'-y_q')} \right)
 \left(\prod_{j=1}^{q}  \int \frac{d^4q_j'}{(2\pi)^4} e^{-iq'_j\cdot (y_j' - y_{j-1}')} \right)
\nonumber \\ \!\!&\times&\!\!
\langle A | \bar{\psi}(y_0) \gamma^\mu \left(\prod_{i=1}^{p} \frac{\slashed{q}_i}{q_i^2 - i\epsilon} \slashed{A}^{a_i}(y_i)\right) \frac{\slashed{q}_{p+1}}{q_{p+1}^2 - i \epsilon} \gamma^\alpha \left(\prod_{i=p+1}^{n} \frac{\slashed{\bar{q}}_i}{\bar{q}_i^2 - i\epsilon} \slashed{A}^{a_i}(y_i)\right) \slashed{l}_q  \nonumber\\&\times&
\left(\prod_{j=m}^{q+1} \slashed{A}^{a_j'}(y_j') \frac{\slashed{\bar{q}}_j'}{\bar{q}_j'^2 + i \epsilon}\right)
 \gamma^\beta \frac{\slashed{q}_{q+1}'}{q_{q+1}'^2 + i \epsilon}
\left(\prod_{j=q}^{1} \slashed{A}^{a_j'}(y'_j) \frac{\slashed{q}_j'}{q_j'^2 + i \epsilon}\right) \gamma^\nu \psi(0) |A\rangle G_{\alpha\beta}(l) .
\end{eqnarray}

\begin{figure}[thb]
\includegraphics[width=1.0\textwidth]{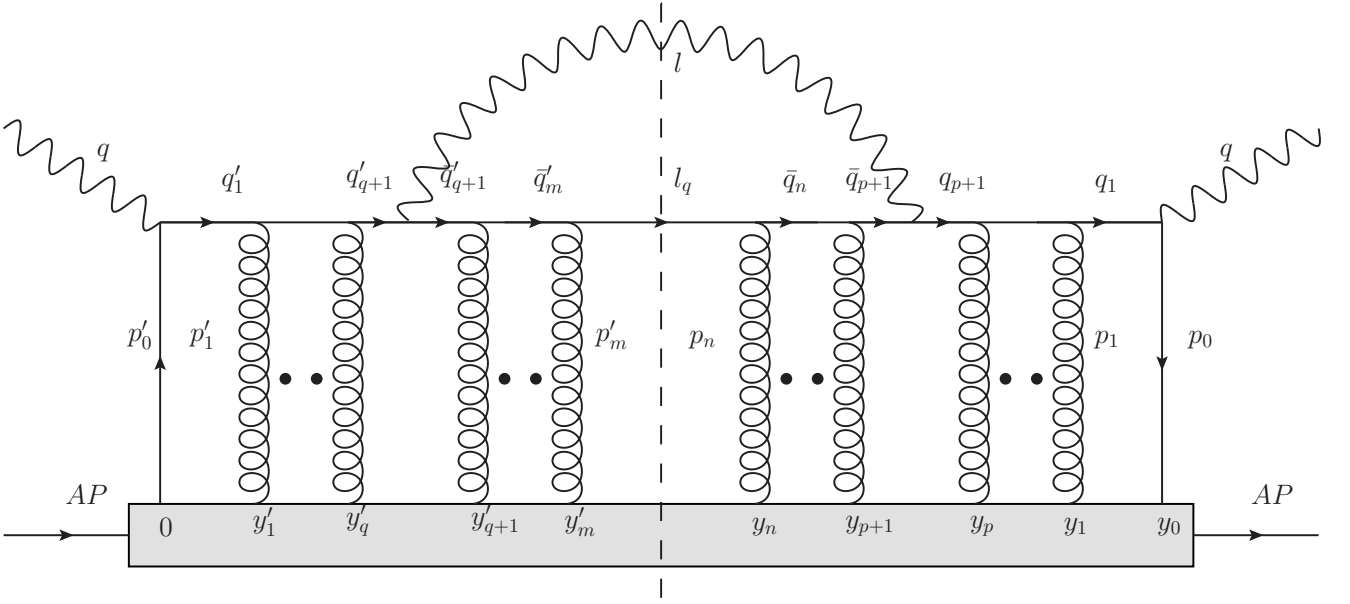}
 \caption{The generic diagram for photon bremsstrahlung process from a quark undergoing multiple scatterings with the medium.
}
\label{mntwist}
\end{figure}

The simplification of the above hadronic tensor is analogous to the case of single scattering presented in the previous section.
We first isolate the phase factor associated with the photon insertion locations $z$ and $z'$: $e^{-i(\bar{q}_{p+1} + l - q_{p+1}) \cdot z}  e^{i(\bar{q}'_{q+1} + l - q'_{q+1}) \cdot z'}$.
The integrations over the locations $z$ and $z'$ can be simply carried out and produce two $\delta$ functions, which we can be used to integrate over the momenta $\bar{q}_{p+1}$ and $\bar{q}'_{q+1}$.
The remaining phase factor may be collected as: $\left(\prod_{i=1}^n e^{-ip_i\cdot y_i} \right) \left(\prod_{j=1}^m e^{ip'_j\cdot y'_j} \right)$.
Using the momentum conservation at each interaction vertex, we may write down the following relations for various momenta shown in Fig. \ref{mntwist}:
\begin{eqnarray}
& q_{i+1} = q + K_i = q + \sum_{j=0}^i p_j,\,\,\,\, (i \le p+1) ;
& \bar{q}_{i+1} = q + K_i - l = q + \sum_{j=0}^i p_j - l, \,\,\,\, (i \ge p+1); \nonumber\\
& q'_{i+1} = q + K'_i = q + \sum_{j=0}^i p'_j, \,\,\,\, (i \le q+1);
& \bar{q}'_{i+1} = q + K'_i -l = q + \sum_{j=0}^i p'_j - l, \,\,\,\, (i \ge q+1). \ \ \ \ \ \
\end{eqnarray}
The new variables $K_i = \sum_{j=0}^i p_i$ and $K'_i = \sum_{j=0}^i p'_i$ represent the total momentum exchanged between the hard quark and the dense nuclear medium.
Using the above relations, one may change the integration variables $q_{i+1} \to p_i$ and $\bar{q}'_{j+1} \to p'_j$.
We may re-introduce the $n$-th momentum $p_n$ by inserting the identity:
\begin{eqnarray}
\int \frac{d^4p_n}{(2\pi)^4} (2\pi)^4 \delta^4(l+l_q - K_{n-1} - p_n - q) = 1.
\end{eqnarray}
Note that the $m$-th momentum $p'_m$ may be obtained as: $p'_m = K_n - K'_{m-1}$.
As mentioned in the previous section, we will perform the calculation in the light-cone gauge ($A^-=0$) in the Breit frame in the limit of very high energy, the dominant component of the vector potential is the forward $(+)$-component.
Since the non-zero contribution involving $A_\perp$ term appears as $(\gamma \cdot q_{i\perp})(\gamma \cdot A_\perp^{a_i})$, which is two-order smaller than that containing $A^+$ term, we may neglect $A_\perp$ component and use the approximation $\slashed{A}= \gamma^- A^+$, i.e, only $(-)$-component survives for these $\gamma$ matrices.
Similar to the previous section, one may also factor out one-nucleon state from the nucleus state and ignore the $(\perp)$-component of the quark field operators.
With the above simplifications, the hadronic tensor now takes the following form:
\begin{eqnarray}
W^{A\mu\nu}_{nmpq} \!\!&=&\!\! \sum_q Q_q^4 e^2 g^{n+m} \frac{1}{N_c} {\rm Tr}\left[\left(\prod_{i=1}^{n} T^{a_i}\right) \left(\prod_{j=m}^{1} T^{a'_j}\right) \right]
\int \frac{d^4l}{(2\pi)^4} (2\pi)\delta(l^2) \int \frac{d^4l_q}{(2\pi)^4} (2\pi)\delta(l_q^2)
\\ \!\!&\times&\!\!
\int d^4y_0 \left(\prod_{i=1}^{n} \int d^4y_i \right) \left(\prod_{j=1}^{m} \int d^4y'_j \right)
\left(\prod_{i=0}^{n-1} \frac{d^4p_i}{(2\pi)^4} \right) \left(\prod_{j=0}^{m-1} \int \frac{d^4p'_j}{(2\pi)^4} \right)
\int \frac{d^4p_n}{(2\pi)^4} (2\pi)^4 \delta^4(l+l_q - K_n - q) \nonumber\\ \!\!&\times&\!\!
\left(\prod_{i=0}^n e^{-ip_i\cdot y_i} \right) \left(\prod_{j=1}^m e^{ip'_j\cdot y'_j} \right)
\left(\prod_{i=1}^{p+1} \frac{1}{q_i^2-i\epsilon} \right) \left(\prod_{i=p+1}^{n} \frac{1}{\bar{q}_i^2-i\epsilon} \right)
\left(\prod_{j=1}^{q+1} \frac{1}{q_j'^2+i\epsilon} \right) \left(\prod_{j=q+1}^{m} \frac{1}{\bar{q}_j'^2+i\epsilon} \right) \nonumber \\ \!\!&\times&\!\!
(-g_\perp^{\mu\nu}) A C_p^A \langle p | \bar{\psi}(y_0) \frac{\gamma^+}{2} \psi(0) |p\rangle
\langle A | \left(\prod_{i=1}^{n} A^{+a_i}(y_i)\right) \left(\prod_{j=m}^{1} A^{+a_j'}(y_j') \right) |A\rangle  \nonumber \\ \!\!&\times&\!\!
{\rm Tr}\left[\frac{\gamma^-}{2} \left(\prod_{i=1}^{p} {\slashed{q}_i} \gamma^- \right) {\slashed{q}_{p+1}} \gamma^\alpha \left(\prod_{i=p+1}^{n} {\slashed{\bar{q}}_i} \gamma^-
\right) \slashed{l}_q  \left(\prod_{j=m}^{q+1} \gamma^- {\slashed{\bar{q}}_j'}\right)
\gamma^\beta {\slashed{q}_{q+1}'} \left(\prod_{j=q}^{1} \gamma^- {\slashed{q}_j'}\right)\right] G_{\alpha\beta}(l). \nonumber \ \
\end{eqnarray}
We now try to isolate the largest length-enhanced contribution which arises when the maximum number of quark lines are close to on-shell.
For this purpose, we look at the quark lines before and after the photon emission,
\begin{eqnarray}
&& q_{i+1}^2 = (q+K_i)^2 = 2p^+q^-(1+ \bar{x}^-_i)[-x_B + \bar{x}_i - \bar{x}_{Di}],
\nonumber\\
&& \bar{q}_{i+1}^2 = (q+K_i-l)^2 = 2p^+q^-(1+ \bar{x}^-_i-y)[-x_B + \bar{x}_i - \bar{x}_{Ci}],
\end{eqnarray}
where the following momentum fraction variables are used again for conveineince,
\begin{eqnarray}
&& \bar{x}_i = \sum_{j=0}^i x_j 
= \frac{K_i^+}{p^+}; \,
\bar{x}^-_i = \sum_{j=0}^i x^-_j 
= \frac{K_i^-}{q^-}; \,
\nonumber\\
&&
\bar{x}_{Di} = \sum_{j=0}^i {x}_{Dj} = \frac{K_{i\perp}^2}{2p^+q^-(1+ \bar{x}^-_i)}; \,
\bar{x}_{Ci} = \sum_{j=0}^i {x}_{Cj} = x_L(1-y) + \frac{( \mathbf{K}_{i\perp} - \mathbf{l}_\perp)^2}{2p^+q^-(1+\bar{x}_i^--y)}.
\end{eqnarray}
Combining the contributions from the denominators of all the internal quark lines with the on-shell condition of the final outgoing quark line, one obtains:
\begin{eqnarray}
D_q \!\!&=&\!\! \frac{C_q }{(2p^+q^-)^{n+m+3}}
\left( \prod_{i=0}^{p} \frac{1}{-x_B+\bar{x}_i-\bar{x}_{Di}} \right)
\left(\prod_{i=p}^{n-1} \frac{1}{-x_B+\bar{x}_i-\bar{x}_{Ci}} \right)
\left( \prod_{j=0}^{q} \frac{1}{-x_B+{\bar{x}'}_i-{\bar{x}'}_{Di}} \right)
 \nonumber\\
\!\!&\times&\!\! \left(\prod_{j=q}^{m-1} \frac{1}{-x_B+{\bar{x}'}_i-{\bar{x}'}_{Ci}} \right)(2\pi)\delta(-x_B+\bar{x}_n-\bar{x}_{Cn}),
\end{eqnarray}
where
\begin{eqnarray}
C_q \!\!&=&\!\! \left( \prod_{i=0}^{p} \frac{1}{1+\bar{x}_i^-} \right) \left(\prod_{i=p}^{n} \frac{1}{1+\bar{x}_i^--y} \right)
\left( \prod_{j=0}^{q} \frac{1}{1+{\bar{x}_j}'^-} \right) \left(\prod_{j=q}^{m-1} \frac{1}{1+{\bar{x}_j}'^--y} \right) .\ \
\end{eqnarray}
Regarding the numerators of the quark lines, one may contract the trace terms with various components of photon polarization sum $G_{\alpha\beta}(l)$.
The terms contracted with the projection $G_{\perp\perp}$ give,
\begin{eqnarray}
N_{\perp\perp} = \frac{2(2q^-)^{n+m+1}}{C_q}
 \left[\frac{\mathbf{K}_{p \perp} \cdot \mathbf{K}_{q\perp}'}{(1+\bar{x}_p^-)(1+{\bar{x}_q}'^-)} + \frac{(\mathbf{K}_{p \perp} -\mathbf{l}_\perp) \cdot (\mathbf{K}_{q\perp}'-\mathbf{l}_\perp)}{(1+\bar{x}_p^--y)(1+{\bar{x}_q}'^--y)}\right].
\end{eqnarray}
Similarly, the terms associated with the projection $G^{++}$ produce,
\begin{eqnarray}
N_{++} = \frac{4(2q^-)^{n+m+1}}{C_q} \frac{l_\perp^2}{y^2}.
\end{eqnarray}
The terms related to the projections $G^{+\perp}$ and $G^{\perp+}$ read,
\begin{eqnarray}
N_{\perp+} + N_{+\perp} = -\frac{2(2q^-)^{n+m+1}}{C_q} \frac{1}{y}
\left[\frac{\mathbf{K}_{p \perp} \cdot \mathbf{l}_{\perp}}{1+\bar{x}_p^-} + \frac{\mathbf{K}_{q \perp}' \cdot \mathbf{l}_{\perp}}{1+{\bar{x}_q}'^-} + \frac{(\mathbf{K}_{p \perp} -\mathbf{l}_\perp) \cdot \mathbf{l}_{\perp}}{1+\bar{x}_p^--y} + \frac{(\mathbf{K}_{q \perp}' -\mathbf{l}_\perp) \cdot \mathbf{l}_{\perp}}{1+{\bar{x}_q}'^--y}\right] . \ \
\end{eqnarray}
With the above simplifications, the hadronic tensor now can be written as the following form:
\begin{eqnarray}
W^{A\mu\nu}_{nmpq} \!\!&=&\!\! \sum_q Q_q^4 e^2 g^{n+m} \frac{1}{N_c} {\rm Tr}\left[\left(\prod_{i=1}^{n} T^{a_i}\right) \left(\prod_{j=m}^{1} T^{a'_j}\right) \right]
\int \frac{d^4l}{(2\pi)^4} (2\pi)\delta(l^2) \int \frac{d^4l_q}{(2\pi)^4}   (2\pi)^4 \delta^4(l+l_q - K_n - q) \nonumber\\\!\!&\times&\!\!
\int d^4y_0 \left(\prod_{i=1}^{n} \int d^4y_i \right) \left(\prod_{j=1}^{m} \int d^4y'_j \right)
\left(\prod_{i=0}^n \int\frac{d^3\mathbf{p}_i dx_i}{(2\pi)^4} \right) \left(\prod_{j=0}^{m-1} \int \frac{d^3\mathbf{p}'_j dx'_j}{(2\pi)^4} \right)
\nonumber\\ \!\!&\times&\!\!
\left(\prod_{i=0}^n e^{-i\mathbf{p}_i\cdot \mathbf{y}_i} e^{-ix_ip^+y_i^-} \right) \left(\prod_{j=0}^m e^{i \mathbf{p}_j'\cdot \mathbf{y}_j' e^{i x_j' p^+ y_i'^-} } \right)
\left( \prod_{i=0}^{p} \frac{1}{-x_B+\bar{x}_i-\bar{x}_{Di}-i\epsilon} \right) \left(\prod_{i=p}^{n-1} \frac{1}{-x_B+\bar{x}_i-\bar{x}_{Ci}-i\epsilon} \right) \nonumber\\\!\!&\times&\!\!
\left( \prod_{j=0}^{q} \frac{1}{-x_B+{\bar{x}'}_i-{\bar{x}'}_{Di}+i\epsilon} \right) \left(\prod_{j=q}^{m-1} \frac{1}{-x_B+{\bar{x}'}_i-{\bar{x}'}_{Ci}+i\epsilon} \right) (2\pi)\delta(-x_B+\bar{x}_n-\bar{x}_{Cn})\nonumber\\
\!\!&\times&\!\!  (-g_\perp^{\mu\nu}) A C_p^A \langle p | \bar{\psi}(y_0) \frac{\gamma^+}{2} \psi(0) |p\rangle
 \langle A | \left(\prod_{i=1}^{n} A^{+a_i}(y_i)\right) \left(\prod_{j=m}^{1} A^{+a'_j}(y'_j) \right) |A\rangle  \nonumber \\ &&
\nonumber\\ \!\!&\times&\!\!
\frac{2}{(2p^+q^-)^2}
\frac{1 + \left(1 - \frac{y}{1+{\bar{x}_p^-}} \right) \left(1 -  \frac{y }{1+{{\bar{x}_q}'^-}} \right)}{y^2\left(1 - \frac{y}{1+{\bar{x}_p^-}} \right) \left(1 -  \frac{y }{1+{{\bar{x}_q}'^-}} \right)}
\left(\mathbf{l}_\perp - \frac{y }{1+{\bar{x}_p^-}} \mathbf{K}_{p\perp} \right)\cdot \left(\mathbf{l}_\perp - \frac{y }{1+{{\bar{x}_q}'^-}} \mathbf{K}_{q\perp}' \right). \ \
\end{eqnarray}
In the above expression, we have changed the integration variables: $p_i^+ \to x_i = p_i^+/p^+$ and ${p'_j}^+ \to x'_j = {p'_j}^+/p^+$.

We now perform the integrations over the momentum fractions $x_i$ and $x'_j$.
Using the overall momentum conservation, $p_m = \sum_{i=0}^n p_i - \sum_{j=0}^{m-1} p'_j$, one may write down the phase factor as follows:
\begin{eqnarray}
\Gamma^+ = \prod_{i=0}^n e^{-ix_ip^+(y_i^--y_m'^-)} \prod_{j=0}^{m-1} e^{ix'_jp^+(y_j'^--y_m'^-)}\!\!,
\end{eqnarray}
Now we may use the on-shell condition for the outgoing quark $(2\pi) \delta(-x_B+\bar{x}_n - \bar{x}_{Cn})$ to integrate out the momentum fraction $x_n$, and re-organize the phase factor as follows:
\begin{eqnarray}
\Gamma^+ = \left(e^{-i(x_B+\bar{x}_{Cn})p^+y_n^-} \prod_{i=0}^{n-1} e^{-ix_ip^+(y_i^--y_n^-)} \right)
\left(e^{i(x_B+\bar{x}_{Cm}')p^+y_m'^-} \prod_{j=0}^{m-1} e^{ix_j' p^+(y_j'^--y_m'^-)} \right) = \Gamma_n^+ \Gamma_m'^+ ,
\end{eqnarray}
where $\Gamma_n^+$ and $\Gamma_m'^+$ denote the parts associated with $x_i$ ($1\le i\le n$) and $x'_j$ ($1\le j\le m$) integrals.
Using the contour integral technique, one may perform all the remaining integrations over $x_i$ and $x'_j$: starting from the propagators adjacent to the cut and proceeding to those adjacent to the photon radiation point.
We first look at the integration over the momentum fraction $x_{n-1}$.
Isolating the corresponding phase factor and closing the contour of $x_{n-1}$ with a counter-clockwise semi-circle in the upper half of the complex plane, one may obtain:
\begin{eqnarray}
\int \frac{dx_{n-1}}{2\pi} \frac{e^{-ix_{n-1}p^+(y_{n-1}^--y_n^-)}}{-x_B + \bar{x}_{n-1} - \bar{x}_{Cn-1} - i \epsilon}  = i \theta(y_n^- - y_{n-1}^-) e^{-i(x_B + \bar{x}_{Cn-1} - \bar{x}_{n-2})p^+(y_{n-1}^- - y_n^-)},
\end{eqnarray}
where the $\theta$-function means that the quark line propagates from the position $y_{n-1}^-$ to the position $y_n^-$.
Combining the above result, the phase factor takes the following form:
\begin{eqnarray}
\Gamma_n^+ \to i\theta(y_n^- - y_{n-1}^-) e^{-i{x}_{Cn}p^+ y_n^-}  e^{-i(x_B+\bar{x}_{Cn-1})p^+y_{n-1}^-}
\left(\prod_{i=0}^{n-2} e^{-ix_ip^+(y_i^--y_n^-)} \right).
\end{eqnarray}
Similarly, one may perform the remaining integrals for the momentum fraction $x_i$ until the one for $x_{p+1}$.
Then the phase factor can be collected as:
\begin{eqnarray}
\Gamma_n^+ \to \left(\prod_{i=p+2}^n i\theta(y_i^- - y_{i-1}^-) e^{-i{x}_{Ci} p^+ y_i^-} \right) e^{-i(x_B+\bar{x}_{Cp+1})p^+y_{p+1}^-}
 \left(\prod_{i=0}^{p} e^{-ix_ip^+(y_i^--y_{p+1}^-)} \right).
\end{eqnarray}
Now performing the contour integration for $x_p$, one may obtain:
\begin{eqnarray}
&& \int \frac{dx_p}{2\pi}  \frac{e^{-ix_p p^+(y_p^--y_{p+1}^-)}}{(-x_B + \bar{x}_p - \bar{x}_{Cp} - i \epsilon)(-x_B + \bar{x}_p - \bar{x}_{Dp} - i \epsilon)} \nonumber\\
&&= {i\theta(y_{p+1}^--y_p^-)   e^{-ix_B p^+(y_p^--y_{p+1}^-)}}
\frac{ e^{-i\bar{x}_{Cp}p^+(y_p^- - y_{p+1}^-)} - e^{-i\bar{x}_{Dp}p^+(y_p^- - y_{p+1}^-)}} {\bar{x}_{Cp} - \bar{x}_{Dp}} e^{i\bar{x}_{p-1}p^+(y_p^--y_{p+1}^-)}. \ \
\end{eqnarray}
Two terms in the above equation come from the fact that either of the two propagators around the photon insertion point can be used to obtain the on-shell conditions for the momentum fraction $x_p$.
Collecting the above result, the phase factor now becomes:
\begin{eqnarray}
\Gamma_n^+ \to \left(\prod_{i=p+1}^n i\theta(y_i^- - y_{i-1}^-) e^{-i{x}_{Ci} p^+ y_i^-} \right) e^{-i(x_B+\bar{x}_{Dp})p^+y_p^-}
\left(\prod_{i=0}^{p-1} e^{-ix_ip^+(y_i^--y_p^-)} \right) \frac{e^{-i\delta\bar{x}_{Dp} p^+ y_p^-} - e^{-i\delta\bar{x}_{Dp} p^+ y_{p+1}^-}} {\delta\bar{x}_{Dp}}.
\end{eqnarray}
The remaining integrations over $x_i$ after the photon insertion may be evaluated in a similar way.
After carrying out the integration over all momentum fractions $x_i$ in the complex conjugate, the phase factor may be cast into the following form:
\begin{eqnarray}
\Gamma_n^+ \to \left(\prod_{i=p+1}^n i\theta(y_i^- - y_{i-1}^-) e^{-i{x}_{Ci} p^+ y_i^-} \right)
\left(\prod_{i=1}^p i\theta(y_i^- - y_{i-1}^-) e^{-i{x}_{Di} p^+ y_i^-} \right)
e^{-ix_Bp^+y_0^-} \frac{e^{-i\delta\bar{x}_{Dp} p^+ y_p^-} - e^{-i\delta\bar{x}_{Dp} p^+ y_{p+1}^-}} {\delta\bar{x}_{Dp}}. \ \
\end{eqnarray}
For the integrations over the momentum fractions $x'_j$ in the amplitude, a close contour of $x_i$ with a clockwise semi-circle in the upper half of the complex plane should be chosen, giving a factor of $(-i)$ instead of $i$ associated with each $\theta$ function.
After the integrations over all the quark lines have been done, the hadronic tensor may be written as:
\begin{eqnarray}
W^{A\mu\nu}_{nmpq}  \!\!&=&\!\! \sum_q Q_q^4 e^2 g^{n+m} \frac{1}{N_c} {\rm Tr}\left[\left(\prod_{i=1}^{n} T^{a_i}\right) \left(\prod_{j=m}^{1} T^{a'_j}\right) \right]
\int \frac{d^4l}{(2\pi)^4} (2\pi)\delta(l^2) \int \frac{d^4l_q}{(2\pi)^4}   (2\pi)^4 \delta^4(l+l_q - K_n - q)
\\\!\!&\times&\!\!
\int d^4y_0 \left(\prod_{i=1}^{n} \int d^4y_i \right) \left(\prod_{j=1}^{m} \int d^4y'_j \right)
\left(\prod_{i=0}^{n} \frac{d^3\mathbf{p}_i}{(2\pi)^3} \right) \left(\prod_{j=0}^{m-1} \int \frac{d^3\mathbf{p}'_j}{(2\pi)^3} \right)
  \left(\prod_{i=0}^n e^{-i\mathbf{p}_i\cdot \mathbf{y}_i} \right) \left(\prod_{j=0}^m e^{i\mathbf{p}_j' \cdot \mathbf{y}_j' } \right) \nonumber\\\!\!&\times&\!\!
e^{-ix_Bp^+y_0^-}  \left(\prod_{i=1}^n i\theta(y_i^- - y_{i-1}^-) e^{-i{x}_{Di} p^+ y_i^-} \right)
\left(\prod_{j=1}^m (-i)\theta(y_j'^- - y_{j-1}'^-) e^{i{x_D}'_j p^+ y_j'^-}\right) \nonumber\\ \!\!&\times&\!\!
(-g_\perp^{\mu\nu}) A C_p^A \langle p | \bar{\psi}(y_0) \frac{\gamma^+}{2} \psi(0) |p\rangle
 \langle A | \left(\prod_{i=1}^{n} A^{+a_i}(y_i)\right) \left(\prod_{j=m}^{1} A^{+a'_j}(y'_j) \right) |A\rangle  \nonumber \\ \!\!&\times&\!\!
 \left(\prod_{i=p+1}^n e^{-i\delta{x}_{Di} p^+ y_i^-} \right)
 \left(\prod_{j=q+1}^n e^{i\delta{x_D}_j' p^+ y_j'^-} \right)
 \left(e^{-i\delta\bar{x}_{Dp} p^+ y_p^-} - e^{-i\delta\bar{x}_{Dp} p^+ y_{p+1}^-}\right)
\left(e^{i\delta\bar{x}_{Dq}' p^+ y_q'^-} - e^{i\delta\bar{x}_{Dq}' p^+ y_{q+1}'^-}\right)
\nonumber\\ \!\!&\times&\!\!
\frac{1}{\delta\bar{x}_{Dp} \delta\bar{x}'_{Dq}}
\frac{2}{(2p^+q^-)^2}
\frac{1 + \left(1 - \frac{y}{1+{\bar{x}_p^-}} \right) \left(1 -  \frac{y }{1+{{\bar{x}_q}'^-}} \right)}{y^2\left(1 - \frac{y}{1+{\bar{x}_p^-}} \right) \left(1 -  \frac{y }{1+{{\bar{x}_q}'^-}} \right)}
\left(\mathbf{l}_\perp - \frac{y }{1+{\bar{x}_p^-}} \mathbf{K}_{p\perp} \right)\cdot \left(\mathbf{l}_\perp - \frac{y }{1+{{\bar{x}_q}'^-}} \mathbf{K}_{q\perp}' \right). \nonumber
\end{eqnarray}

We now perform the resummation over different photon insertions locations ($y_p$ in the complex conjugate and $y_q$ in the amplitude) for given numbers of scatterings experienced by the propagating hard quark jet.
In this work, we only consider the case with the same number of scatterings in both the amplitude and the complex conjugate ($n=m$).
The diagrams with $n\ne m$ contribute to the gauge corrections to the terms with $\min(n,m)$ scatterings experienced by the hard quark jet \cite{Majumder:2007hx, Majumder:2007ne, Qin:2012fua}.
Similar to the previous section, the following assumption is invoked to simplify the matrix elements of $2n (n=m)$ gluon vector potentials in the nuclear state (note that the quark operators has already been factorized out).
In the very high energy limit, nucleons are traveling in straight lines and are almost independent of each other over the time period of the interaction between the hard quark jet and the nuclear medium.
Thus it is reasonable to approximate the nucleus by a weakly-interacting homogenous gas of nucleons, i.e., the expectation of gluon field operators in the nucleus states may be cast into a product of the expectations in the nucleon states.
By noting that a nucleon is a color singlet, the combination of gluon (or quark) field insertions must be restricted to a color singlet, i.e., the first non-zero and the largest contribution are the terms with $2n$ gluon insertions divided into $n$ singlet pairs in $n$ separate nucleon states (note that the atom number $A$ of the nucleus is assumed to be very large).
The above statement means that the expectation of the $2n$ gluon operators in the nucleus state can be decomposed as follows:
\begin{eqnarray}
 \langle A | \left(\prod_{i=1}^{n} A^{+a_i}(y_i)\right) \left(\prod_{j=n}^{1} A^{+a'_j}(y'_j) \right) |A\rangle
= \left(\frac{\rho}{2p^+}\right)^n \left(\prod_{i=1}^{n} \frac{\delta_{a_i a'_i}}{N_c^2 - 1} \langle p| A^+(y_i) A^+(y'_i) | p\rangle\right),
\end{eqnarray}
where the colors of gluon field operators have been averaged.
The trace for the color matrices can be easily evaluated and gives:
\begin{eqnarray}
\frac{1}{N_c(N_c^2 - 1)^n} {\rm Tr}\left[\left(\prod_{i=1}^{n} T^{a_i}\!\right) \!\! \left(\prod_{i=n}^1 T^{a'_i}\!\right) \right] = \left( \frac{C_F}{N_c^2 - 1}\right)^n.
\end{eqnarray}
Now for convenience, we change the four-vector location variables: $(y_i,y_i') \to (Y_i, \delta y_i)$,
\begin{eqnarray}
{Y}_i = \frac{1}{2}({y}_i + {y}'_i);  &&  \delta{y}_i = {y}_i - {y}'_i.
\end{eqnarray}
Using the translational invariance of the correlation functions, $\langle p| A^+(y_i) A^+(y'_i) | p\rangle \approx \langle p| A^+(\delta y_i^-, \delta \mathbf{y}_i^-) A^+(0) | p\rangle$, the integration for the phase factor may be performed:
\begin{eqnarray}
\int d^3\mathbf{y}_i \int d^3\mathbf{y}'_i e^{-i\mathbf{p}_i \cdot \mathbf{y}_i} e^{i\mathbf{p}'_i\cdot \mathbf{y}'_i}
= (2\pi)^3 \delta^3(\mathbf{p}_i - \mathbf{p}'_i) \int d^3\delta \mathbf{y}_i e^{-i(\mathbf{p}_i+\mathbf{p}_i')\cdot \frac{\delta\mathbf{y}_i}{2}}.
\end{eqnarray}
In the above expression, the $\delta$ function represents the fact that each pair of gluon field insertions in each nucleon states carry the same momentum, $\mathbf{p}_i' = \mathbf{p}_i$. 
The $\delta$ functions may be utilized to carry out the integration over the momentum $\mathbf{p}'_i$.
Recalling the expressions of the momentum fraction variables $\bar{x}_{Dp}$ and $\bar{x}_{Cp}$, one may obtain the momentum fraction $\delta \bar{x}_{Dp}$:
\begin{eqnarray}
\delta\bar{x}_{Dp} \!\!&=&\!\! \bar{x}_{Cp} - \bar{x}_{Dp}
= \frac{\left(\mathbf{l}_\perp - \frac{y }{1+\bar{x}_p^-} \mathbf{K}_{p\perp} \right)^2}{2p^+q^-y\left(1-\frac{y}{1+ \bar{x}_p^-}\right)}.
\end{eqnarray}
The expression of $\delta \bar{x}'_{Dq}$ is completely analogous.
Using the above expressions, one may obtain the hard part of the matrix element (the last line of the hadronic tensor, denoted as $T_{pq}$) as follows:
\begin{eqnarray}
\label{linear_approximation}
T_{pq} \!\!&=&\!\! \frac{2yP(y)}{l_\perp^2}
\left[1 + \frac{y(1-y)}{1+(1-y)^2} \frac{K_p^-+{K'_q}^-}{q^-}
+ y \frac{\mathbf{l}_\perp \cdot (\mathbf{K}_{p\perp}+\mathbf{K}'_{q\perp})}{l_\perp^2}
-\frac{y (1-y)}{1+(1-y)^2} \frac{{K_p^-}^2+{K_q'^-}^2}{{q^-}^2}
\right. \\ \!\!&&\!\!
+\frac{y^2}{1+(1-y)^2}\frac{K_p^-\cdot{K'_q}^-}{{q^-}^2}
-y^2\frac{K_{p\perp}^2+K_{q\perp}'^2}{l_\perp^2} + 2y^2\frac{(\mathbf{l}_\perp \cdot \mathbf{K}_{p\perp})^2+(\mathbf{l}_\perp \cdot \mathbf{K}'_{q\perp})^2}{l_\perp^4}
+y^2\frac{\mathbf{K}_{p\perp}\cdot\mathbf{K}'_{q\perp}}{l_\perp^2}
\nonumber \\  \!\!&&\!\!
\left.
+\left(\frac{y^2(1-y)}{1+(1-y)^2}-y\right)\left(\frac{\mathbf{l}_\perp \cdot \mathbf{K}_{p\perp}}{l_\perp^2}\frac{K_p^-}{q^-}
+\frac{\mathbf{l}_\perp \cdot \mathbf{K}'_{q\perp}}{l_\perp^2} \frac{{K'_q}^-}{q^-} \right)
+\frac{y^2(1-y)}{1+(1-y)^2} \left(\frac{\mathbf{l}_\perp \cdot \mathbf{K}_{p\perp}}{l_\perp^2}\frac{{K'_q}^-}{q^-}
+\frac{\mathbf{l}_\perp \cdot \mathbf{K}'_{q\perp}}{l_\perp^2} \frac{K_p^-}{q^-} \right)\right]. \nonumber
\end{eqnarray}
In the above expression, we have kept the terms up to the second order in $\frac{K_i^-}{q^-}$ and $\frac{{K}_{i\perp}}{{l}_\perp}$ and their cross terms, to be consistent with the momentum gradient expansion up to the second order as will be done in a short moment.
Now we simplify the phase factor (the second line of the hadronic tensor). Keeping the leading contribution, 
the phase factor (denoted as $S_{pq}$) can be obtained as follows:
\begin{eqnarray}
S_{pq} = \left({e^{-ix_L p^+ y_p^-} - e^{-ix_L p^+ y_{p+1}^-}}\right)
\left({e^{ix_L p^+ {y'_q}^-} - e^{ix_L p^+ {y'}_{q+1}^-}}\right)
\end{eqnarray}
With the above simplifications, the sum over the photon insertion locations $y_p$ and $y_q'$ may be performed:
\begin{eqnarray}
\sum_{p=0}^{n} \sum_{q=0}^n S_{pq} T_{pq} \!\!&=&\!\! \frac{2yP(y)}{l_\perp^2} e^{-ix_Lp^+y_0^-} \left[1 + \frac{y(1-y)}{1+(1-y)^2} \sum_{k=1}^n \frac{p_k^-}{q^-} F_k + y \sum_{k=1}^n \frac{\mathbf{l}_\perp \cdot \mathbf{p}_{k\perp}}{l_\perp^2} F_k
\nonumber\right. \\ \!\!&&\!\!
- \frac{y(1-y)}{1+(1-y)^2}\sum_{k=1}^n\frac{{K_k^-}^2-(K_{k-1}^-)^2}{{q^-}^2}F_k
+\frac{y^2}{1+(1-y)^2}\sum_{k=1}^n\sum_{p=1}^n \frac{p_k^- p_p^-}{{q^-}^2} G_{kp}
\nonumber\\ \!\!&&\!\!
-y^2\sum_{k=1}^n\frac{K_{k\perp}^2-K_{k-1 \perp}^2}{l_\perp^2}F_k
+2y^2 \sum_{k=1}^n \frac{(\mathbf{l}_\perp \cdot \mathbf{K}_{k\perp})^2-(\mathbf{l}_\perp \cdot \mathbf{K}_{{k-1}\perp})^2}{{l_{\perp}^4}}F_k
\nonumber\\ \!\!&&\!\!
+\left(\frac{y^2(1-y)}{1+(1-y)^2}-y\right) \sum_{k=1}^n \left(\frac{\mathbf{l}_\perp \cdot \mathbf{K}_{k\perp}}{l_\perp^2} \frac{K_k^-}{q^-}-\frac{\mathbf{l}_\perp \cdot \mathbf{K}_{{k-1}\perp}}{l_\perp^2}\frac{K_{k-1}^-}{q^-}\right)F_k
\nonumber\\ \!\!&&\!\!
\left.
+y^2\sum_{k=1}^n\sum_{p=1}^n \frac{\mathbf{p}_{k\perp}\cdot \mathbf{p}_{p\perp}}{{q^-}^2} G_{kp}+\frac{2y^2(1-y)}{1+(1-y)^2}\sum_{k=1}^n\sum_{p=1}^n\frac{\mathbf{l}_\perp \cdot \mathbf{p}_{p\perp}}{l_\perp^2} \frac{p_k^-}{q^-} G_{kp}\right].
\end{eqnarray}
where
\begin{eqnarray}
&& F_k = 
{e^{-ix_L p^+ (Y_k^-+\frac{1}{2}\delta y_k^- - y_0^-)} + e^{ix_L p^+ (Y_k^- - \frac{1}{2}\delta y_k^-)}};
\\
&& G_{kp} = 
{e^{-ix_L p^+ (Y_k^-+\frac{1}{2}\delta y_k^- - y_0^-)}  e^{ix_L p^+ (Y_p^- - \frac{1}{2}\delta y_p^-)}} .
\end{eqnarray}
Note that $Y_k^- (Y_p^-)$ is the location of the photon insertion point and can span over the nucleus size, while $y_0^-$ and $\delta y_k^- (\delta y_p^-)$ are confined within the nucleon size.
Therefore, $y_0^-$ and $\delta y_k^- (\delta y_p^-)$ are small as compared to $Y_k^-(Y_p^-)$, and can be safely dropped in the expressions of $F_k$ and $G_{kp}$. Then we obtain:
\begin{eqnarray}
F_k \approx  2 \cos(x_L p^+ Y_k^-), \, G_{kp} \approx \cos(x_L p^+ ({Y_k}^--{Y_p}^-)).
\end{eqnarray}
where we have exchanged the subscript $k$ and $p$ to symmetrize the expression of $G_{kp}$.
After carrying out the sum over the photon insertion locations, the hadronic tensor takes the following form:
\begin{eqnarray}
\label{W_Amunu}
W^{A\mu\nu}_{nn} \!\!&=&\!\! \sum_{p}\sum_{q} W^{A\mu\nu}_{nnpq} = (-g_\perp^{\mu\nu}) A C_p^A \sum_q Q_q^4 \frac{\alpha_e}{2\pi}
\int dy P(y) \int \frac{d^2l_\perp}{\pi l_\perp^2} \int{d^3\mathbf{l}_q}
\int dy_0^- e^{-i(x_B+x_L)p^+y_0^-}  \langle p | \bar{\psi}(y_0^-) \frac{\gamma^+}{2} \psi(0) |p\rangle \nonumber\\\!\!&\times&\!\!
\frac{1}{n!} \left(\prod_{i=1}^{n} \int dY_i^- \int d\delta y_i^- \left(g^2\frac{C_F}{N_c^2 - 1}\frac{\rho}{2p^+}\right) \int d^3\delta \mathbf{y}_i \int \frac{d^3\mathbf{p}_i}{(2\pi)^3} e^{-i\mathbf{p}_i\cdot \delta \mathbf{y}_i} \langle p | A^+(\delta y_i^-, \delta \mathbf{y}_i) A^+(0)  |p\rangle \right)
\nonumber \\ \!\!&\times&\!\!
\left[1 + \frac{y(1-y)}{1+(1-y)^2} \sum_{k=1}^n \frac{p_k^-}{q^-} F_k + y \sum_{k=1}^n \frac{\mathbf{l}_\perp \cdot \mathbf{p}_{k\perp}}{l_\perp^2} F_k - \frac{y(1-y)}{1+(1-y)^2}\sum_{k=1}^n\frac{{K_k^-}^2-(K_{k-1}^-)^2}{{q^-}^2}F_k
\nonumber\right. \\ \!\!&&\!\!
+\frac{y^2}{1+(1-y)^2}\sum_{k=1}^n\sum_{p=1}^n \frac{p_k^- p_p^-}{{q^-}^2} G_{kp}-y^2\sum_{k=1}^n\frac{K_{k\perp}^2-K_{{k-1}\perp}^2}{l_\perp^2}F_k+2y^2 \sum_{k=1}^n \frac{(\mathbf{l}_\perp \cdot \mathbf{K}_{k\perp})^2-(\mathbf{l}_\perp \cdot \mathbf{K}_{{k-1}\perp})^2}{{l_{\perp}^4}}F_k
\nonumber\\ \!\!&&\!\!
+y^2\sum_{k=1}^n\sum_{p=1}^n \frac{\mathbf{p}_{k\perp}\cdot \mathbf{p}_{p\perp}}{{q^-}^2} G_{kp}+\left(\frac{y^2(1-y)}{1+(1-y)^2}-y\right) \sum_{k=1}^n \left(\frac{\mathbf{l}_\perp \cdot \mathbf{K}_{k\perp}}{l_\perp^2} \frac{K_k^-}{q^-}-\frac{\mathbf{l}_\perp \cdot \mathbf{K}_{{k-1}\perp}}{l_\perp^2}\frac{K_{k-1}^-}{q^-}\right)F_k
\nonumber\\ \!\!&&\!\!
\left.+\frac{2y^2(1-y)}{1+(1-y)^2}\sum_{k=1}^n\sum_{p=1}^n\frac{\mathbf{l}_\perp \cdot \mathbf{p}_{p\perp}}{l_\perp^2} \frac{p_k^-}{q^-} G_{kp}\right] \delta^3(\mathbf{l}+\mathbf{l}_q - \mathbf{K}_n - \mathbf{q}).
\end{eqnarray}
One can clearly see in the above expression many terms originating from the coupling between different scatterings, which are absent for the case of single scattering.

We now perform the resummation over the number of multiple scatterings experienced by the hard quark jet.
Similar to the previous section, we introduce the momentum gradient expansion for the hard part $H(\mathbf{p}_i)$ of the matrix elements (the last four lines in the above hadronic tensor) in order to obtain a closed formula for single photon radiation spectrum.
Assuming the momentum exchange in each of the multiple scatterings between the hard parton and the medium constituents is small, one may expand the hard part $H(\mathbf{p}_i)$ as a series of the Taylor expansion in the three-dimensional momenta $\mathbf{p}_i = (p_i^-, \mathbf{p}_{i\perp})$ around $\mathbf{p}_i \to 0$:
\begin{eqnarray}
\label{gradient_expansion_n}
H = \left(\prod_{i=1}^{n} \left[1 + p_i^\alpha \frac{\partial}{\partial p_i^\alpha} + \frac{1}{2} p_i^\alpha p_i^\beta \frac{\partial}{\partial p_i^\alpha} \frac{\partial}{\partial p_i^\beta} + \cdots \right] \right) H|_{\mathbf{p}_1 \cdots \mathbf{p}_n = 0}. \ \
\end{eqnarray}
where $\alpha$ and $\beta$ take the values of ``$-$'' or ``$\perp$".
In the above expression, we have only kept the expansion up to the second order derivative in both longitudinal and transverse momenta.
High-order terms are neglected and should be straightforward to include in the gradient expansion; they correspond to higher-order moments for the momentum distribution of the exchanged gluons.
The zeroth order term (with on momentum derivative) represent the case where the exchanged gluons carry zero momenta; they contribute to the gauge corrections to the diagrams with lower number of gluon insertions which carry nonzero momenta \cite{Luo:1992fz, Luo:1994np, Majumder:2007hx, Majumder:2007ne}, and will not be considered further.
Now performing the integration over $p_i$ by part, one may obtain:
\begin{eqnarray}
&&\langle p| A^+(\delta \mathbf{y}_i) A^+(0)|p \rangle e^{-\mathbf{p}_i \cdot \delta\mathbf{y}_i} p_i^\alpha \frac{\partial}{\partial p_i^\alpha} =   e^{-\mathbf{p}_i \cdot \delta\mathbf{y}_i} (-i) \langle p| \partial^\alpha A^+(\delta \mathbf{y}_i) A^+(0)|p \rangle \frac{\partial}{\partial p_i^\alpha},  \\
&&\langle p| A^+(\delta \mathbf{y}_i) A^+(0)|p \rangle e^{-\mathbf{p}_i \cdot \delta\mathbf{y}_i} p_i^\alpha p_i^\beta \frac{\partial}{\partial p_i^\alpha} \frac{\partial}{\partial p_i^\beta} = e^{-\mathbf{p}_i \cdot \delta\mathbf{y}_i} \langle p| \partial^\alpha A^+(\delta \mathbf{y}_i) \partial^\beta A^+(0)|p \rangle \frac{\partial}{\partial p_i^\alpha} \frac{\partial}{\partial p_i^\beta}.
\end{eqnarray}
We note the condition $\mathbf{p}_i \to 0$ imposed by the Taylor expansion.
This means that the hard part $H$ no longer depends on $\mathbf{p}_i$, thus the integrations for $\mathbf{p}_i$ and $\delta\mathbf{y}_i$ may be carried out directly.
Now we may perform the resummation of the number of multiple scatterings and obtain the hadronic tensor as follows:
\begin{eqnarray}
W^{A\mu\nu} = \sum_n W^{A\mu\nu}_{nn} = (-g_\perp^{\mu\nu}) A C_p^A \sum_q Q_q^4 \frac{\alpha_e}{2\pi}
\int dy P(y) \int \frac{d^2l_\perp}{\pi l_\perp^2} \int{d^3\mathbf{l}_q} (2\pi) f_q(x_B+x_L)\phi(L^-, l_q^-, \mathbf{l}_{q\perp}), \ \
\end{eqnarray}
where
\begin{eqnarray}
\phi(L^-, l_q^-, \mathbf{l}_{q\perp}) = \sum_{n=0}^{\infty} \frac{1}{n!} \left(\prod_{i=1}^{n} \int dY_i^- \left[ - D_{L1}\frac{\partial}{\partial p_i^-} + \frac{1}{2} D_{L2} \frac{\partial^2}{\partial^2 p_i^-}  + \frac{1}{2} D_{T2} {\nabla_{p_{i \perp}}^2}  \right] \right) H|_{\mathbf{p}_1 \cdots \mathbf{p}_n = 0}. \ \
\end{eqnarray}
The transport coefficients $D_{L1}$, $D_{L2}$ and $D_{T2}$ have been defined in the previous section.
Now we look at ten terms in the hard part one by one:
\begin{eqnarray}
H = H_0 + H_-^{(1)} + H_\perp^{(1)} + H_{-, 1}^{(2)} + H_{-, 2}^{(2)} + H_{\perp, 1}^{(2)} +H_{\perp, 2}^{(2)} + H_{\perp, 3}^{(2)} + H_{- \perp, 1}^{(2)} + H_{- \perp, 2}^{(2)}.
\end{eqnarray}
Accordingly, the distribution function $\phi$ may be splitted into ten contributions:
\begin{eqnarray}
\phi =\phi^{(0)} +\phi_-^{(1)} + \phi_\perp^{(1)} + \phi_{-,1}^{(2)} + \phi_{-, 2}^{(2)} + \phi_{\perp, 1}^{(2)} +\phi_{\perp, 2}^{(2)} + \phi_{\perp, 3}^{(2)} + \phi_{- \perp, 1}^{(2)} + \phi_{- \perp, 2}^{(2)}.
\end{eqnarray}
Resumming the number of multiple scatterings, one may obtain the contribution from the term $H_0$ to $\phi^{(0)}$ as follows:
\begin{eqnarray}
\phi^{(0)}  = \exp\left( L^- \left[ D_{L1}\frac{\partial}{\partial l_q^-} + \frac{1}{2} D_{L2} \frac{\partial^2}{\partial^2 l_q^-} + \frac{1}{2} D_{T2} {\nabla_{l_{q\perp}}^2}  \right] \right) \delta(l_q^- - q^-(1-y)) \delta^2(\mathbf{l}_{q\perp} + \mathbf{l}_{\perp}),
\end{eqnarray}
where the derivatives over $p_i$ have been converted to the derivatives over $l_q$.
Obviously, the distribution function $\phi^{(0)}$ satisfies the following differential equation:
\begin{eqnarray}
\frac{\partial \phi^{(0)}}{\partial L^-} = \left[ D_{L1}\frac{\partial}{\partial l_q^-} + \frac{1}{2} D_{L2} \frac{\partial^2}{\partial^2 l_q^-} + \frac{1}{2} D_{T2} {\nabla_{l_{q\perp}}^2}  \right] \phi^{(0)}(L^-, l_q^-, \mathbf{l}_{q\perp}).
\end{eqnarray}
The above equation describes the time evolution of the three-dimensional momentum distribution of a propagating hard parton which only experiences multiple scatterings with the medium constituents but without any radiation \cite{Qin:2012fua}.
Three terms in the above equation represent the contributions from the longitudinal momentum loss and diffusion, and the diffusion of transverse momentum.
If there is no radiation, the initial condition takes the form: $\phi^{(0)}(L^- = 0, l_q^-, \mathbf{l}_{q\perp}) =  \delta(l_q^- - q^-) \delta^2(\mathbf{l}_{q\perp} + \mathbf{l}_{\perp})$, and we return the solution obtained in Ref. \cite{Qin:2012fua}.
Here we consider the photon bremsstrahlung process, whose initial condition is $\phi^{(0)}(L^- = 0, l_q^-, \mathbf{l}_{q\perp}) = \delta(l_q^- - q^-(1-y)) \delta^2(\mathbf{l}_{q\perp} + \mathbf{l}_{\perp})$. Using such initial condition, the distribution function $\phi^{(0)}$ takes the following solution:
\begin{eqnarray}
\phi^{(0)} = \frac{e^{-{(l_q^- - q^-(1-y) + D_{L1}L^-)^2}/({2D_{L2}L^-})}}{\sqrt{2\pi D_{L2} L^-}}   \frac{e^{-{(\mathbf{l}_{q\perp} + \mathbf{l}_{\perp})^2}/({2D_{T2}L^-})}}{2\pi D_{T2} L^-}.
\end{eqnarray}
This solution means that the parton radiates photon immediately after the initial hard collisions, and then experiences multiple scatterings in the nuclear medium.
Now we look at the second term which reads as follows:
\begin{eqnarray}
\phi_-^{(1)} \!\!&=&\!\!  \frac{y(1-y)}{1+(1-y)^2} \sum_{n=1}^{\infty}  \frac{1}{n!}  \sum_{k=1}^n \int dY_k^- F_k \frac{1}{q^-} \left(-D_{L1} - D_{L2} \frac{\partial}{\partial l_q^-} \right)
\nonumber\\ \!\!&\times&\!\!
\left( \prod_{i=1, i\ne k}^n \int dY_i^- \left(D_{L1} \frac{\partial}{\partial l_q^-} + \frac{1}{2} D_{L2} \frac{\partial^2}{\partial^2 l_q^-} + \frac{1}{2} D_{T2} {\nabla_{l_{q\perp}}^2}   \right) \right)
\delta(l_q^- - q^-(1-y)) \delta^2(\mathbf{l}_{q\perp} + \mathbf{l}_{\perp}).
\end{eqnarray}
Considering a homogenous medium, i.e., the transport coefficients $D_{L1}$, $D_{L2}$ and $D_{T2}$ are position independent, then the $Y_k^-$ integral for the function $F_k$ can be carried out directly,
\begin{eqnarray}
\int_0^{L^-} dY_k^- F_k \approx \int_0^L dY_k^- 2 \cos(x_L p^+ Y_k^-) = \frac{2\sin(x_L p^+ L^-)}{x_L p^+} = F_L L^-.
\end{eqnarray}
The distribution function $\phi_-^{(1)}(L^-, l_q^-, \mathbf{l}_{q\perp})$ can be related to $\phi^{(0)}$ as follows:
\begin{eqnarray}
\phi_-^{(1)}  = \frac{y(1-y)}{1+(1-y)^2} F_L \left(-\frac{D_{L1} L^-}{q^-}  - \frac{D_{L2} L^-}{q^-}\frac{\partial}{\partial l_q^-} \right) \phi^{(0)}.
\end{eqnarray}
The analysis of other eight terms are completely analogous, and they read as follows:
\begin{eqnarray}
&& \phi_\perp^{(1)} = y F_L  \frac{ D_{T2}L^-}{l_\perp^2} (-\mathbf{l}_\perp \cdot \nabla_{l_{q\perp}})\phi^{(0)};
\nonumber\\
&&\phi_{-,1}^{(2)}= -\frac{y(1-y)}{1+(1-y)^2} F_L   \left( \frac{D_{L2} L^-}{{{q^-}^2}}+ \left(-\frac{D_{L1}L^-}{{{q^-}}} - \frac{D_{L2}L^-}{{{q^-}}} \frac{\partial}{\partial l_q^-} \right)^2 \right)\phi^{(0)};
\nonumber\\
&& \phi_{-,2}^{(2)} = \frac{y^2}{1+(1-y)^2}  \left(\frac{D_{L2} L^-}{{{q^-}^2}} + G_L \left(-\frac{D_{L1}L^-}{{{q^-}}} - \frac{D_{L2}L^-}{{{q^-}}} \frac{\partial}{\partial l_q^-} \right)^2 \right)\phi^{(0)};
\nonumber\\
&&\phi_{\perp, 1}^{(2)} = -y^2 F_L \left(\frac{2 D_{T2}L^-}{{l_\perp^2}} + \frac{D_{T2}^2 {L^-}^2}{{l_\perp^4}} {\nabla_{l_{q\perp}}^2} \right)\phi^{(0)};
\nonumber\\
&&\phi_{\perp, 2}^{(2)} = 2y^2 F_L\left(\frac{D_{T2}L^-}{l_\perp^2}+ \frac{D_{T2}^2{L^-}^2}{l_\perp^4}(\mathbf{l}_\perp \cdot \nabla_{l_{q\perp}})^2 \right)\phi^{(0)};
\nonumber\\
&&\phi_{\perp, 3}^{(2)} = y^2  \left(\frac{2D_{T2} L^-}{l_\perp^2} + F_L \frac{D_{T2}^2 {L^-}^2}{l_\perp^2} \nabla_{l_{q\perp}}^2 \right)\phi^{(0)};
\nonumber\\
&&\phi_{- \perp, 1}^{(2)} = \left(\frac{y^2(1-y)}{1+(1-y)^2}-y\right) F_L  \left(-\frac{D_{L1}L^-}{q^-} - \frac{D_{L2}L^-}{q^-} \frac{\partial}{\partial l_q^-} \right)\frac{D_{T2}L^-}{l_\perp^2} (-\mathbf{l}_\perp \cdot \nabla_{l_{q\perp}})\phi^{(0)};
\nonumber\\
&&\phi_{- \perp, 2}^{(2)} = \frac{2y^2(1-y)}{1+(1-y)^2} G_L   \left(-\frac{D_{L1}L^-}{q^-} - \frac{D_{L2}L^-}{q^-} \frac{\partial}{\partial l_q^-} \right)\frac{D_{T2}L^-}{l_\perp^2} (-\mathbf{l}_\perp \cdot \nabla_{l_{q\perp}})\phi^{(0)};
\end{eqnarray}
where
\begin{eqnarray}
G_L = \frac{2-2\cos(x_Lp^+L^-)}{(x_Lp^+L^-)^2}
\end{eqnarray}
Putting the above ten contributions together, the hadronic tensor takes the follow form:
\begin{eqnarray}
W^{A \mu\nu} \!\!&=&\!\!  (-g_\perp^{\mu\nu}) A C_p^A \sum_q Q_q^4 \frac{\alpha_e}{2\pi}
\int dy P(y)\int \frac{d^2 l_\perp}{\pi l_\perp^2}\int d^3 \mathbf{l}_q (2\pi) f_q(x_B + x_L)
\\
\!\!&\times&\!\!
 \left\{ 1 + \frac{y(1-y)}{1+(1-y)^2} F_L \left(-\frac{D_{L1}L^-}{q^-} - \frac{D_{L2}L^-}{q^-} \frac{\partial}{\partial l_q^-} \right) + y F_L \frac{ D_{T2}L^-}{l_\perp^2} (-\mathbf{l}_\perp \cdot \nabla_{l_{q\perp}})
\nonumber\right.\\ \!\!&&\!\!
+\frac{y}{1+(1-y)^2} \left[y G_L - (1-y)F_L \right] \left(-\frac{D_{L1}L^-}{{{q^-}}} - \frac{D_{L2}L^-}{{{q^-}}} \frac{\partial}{\partial l_q^-} \right)^2
+\frac{y}{1+(1-y)^2} \left[y-(1-y)F_L\right] \frac{D_{L2}L^-}{{q^-}^2}
\nonumber\\ \!\!&&\!\!
+2y^2\frac{D_{T2}L^-}{l_\perp^2}
+y^2(G_L - F_L)\frac{D_{T2}^2{L^-}^2  }{l_\perp^2}\nabla_{l_{q\perp}}^2
+2y^2F_L\frac{D_{T2}^2{L^-}^2}{l_\perp^4}(\mathbf{l}_\perp \cdot \nabla_{l_{q\perp}})^2
\nonumber\\ \!\!&&\!\!
\left.
+\left[\frac{y^2(1-y)}{1+(1-y)^2} (F_L +2G_L) - y F_L\right]
\left(-\frac{D_{L1}L^-}{q^-} - \frac{D_{L2}L^-}{q^-} \frac{\partial}{\partial l_q^-} \right)\frac{D_{T2}L^-}{l_\perp^2} (-\mathbf{l}_\perp \cdot \nabla_{l_{q\perp}})\right\}
\phi^{(0)}(L^-, l_q^-, \mathbf{l}_{q\perp}).
\nonumber
\end{eqnarray}
As has been mentioned, the first term in the above equation denotes the scenarios in which the photon is radiated immediately after the initial hard scattering, and the subsequent multiple scatterings on the propagating parton is encoded by $\phi^{(0)}$.
The other terms describes the effect that the propagating parton experiences multiple scatterings with the medium constituents, which induces a photon radiation.
In the high energy and collinear limits with small momentum exchange between the hard parton and the medium, one can integrate out the three-dimensional momentum of the final outgoing quark $\mathbf{l}_q=(l_q^-,\mathbf{l}_{q\perp})$, and obtain the single-differential hadronic tensor as follows:
\begin{eqnarray}
\frac{dW^{A \mu\nu}}{dy d{l^2_{\perp}}} \!\!&=&\!\!  (-g_\perp^{\mu\nu}) A C_p^A \sum_q Q_q^4 \frac{\alpha_e}{2\pi}\frac{P(y)}{l_\perp^2} (2\pi) f_q(x_B + x_L)
\left\{ 1 - \frac{y(1-y)}{1+(1-y)^2} F_L \frac{D_{L1}L^-}{q^-}
\right.\\\!\!&&\!\!
\left.
+\frac{y}{1+(1-y)^2} \left[y G_L-(1-y) F_L\right]\left(\frac{D_{L1}L^-}{q^-}  \right)^2
+\frac{y}{1+(1-y)^2} \left[y-(1-y)F_L\right] \frac{D_{L2}L^-}{{q^-}^2}
+2 y^2\frac{D_{T2}L^-}{l_\perp^2}
\right\}.
\nonumber
\end{eqnarray}
From the above expression, one may read off the medium-induced photon bremsstrahlung spectrum:
\begin{eqnarray}
\label{dNgamma_nn}
\frac{dN_\gamma^{\rm med}}{dy d{l^2_{\perp}}} \!\!&=&\!\!  \frac{\alpha_e}{2\pi}\frac{P(y)}{l_\perp^2}
\left\{ - \frac{y(1-y)}{1+(1-y)^2} F_L \frac{D_{L1}L^-}{q^-}
+\frac{y}{1+(1-y)^2} \left[y-(1-y)F_L\right] \frac{D_{L2}L^-}{{q^-}^2}
+2 y^2\frac{D_{T2}L^-}{l_\perp^2}
\nonumber\right.\\\!\!&&\!\!
\left.
+\frac{y}{1+(1-y)^2} \left[y G_L-(1-y) F_L \right]\left(\frac{D_{L1}L^-}{q^-}  \right)^2
\right\}.
\end{eqnarray}
\end{widetext}
One can clearly see from the above equation the individual contributions from the longitudinal drag and the diffusions of longitudinal and transverse momentum transfers to the single photon bremsstrahlung spectrum.
Comparing to the single scattering result shown in Eq. (\ref{dNgamma_11}), an additional contribution from the longitudinal drag (the last term) is obtained for the case of multiple scatterings.
This term originates from the coupling between different scatterings as can be seen in Eq. (\ref{W_Amunu}).
Since we only keep the terms up to the second order in the momentum gradient expansion, only the term quadratic in the drag coefficients $D_{L1}$ survives.
One would expect more additional terms to appear for multiple scattering scenario if higher-order terms are kept in the momentum gradient expansion.
We can see that the additional term goes as $(D_{L1} L^-/q^-)^2$, therefore, it should provide a subleading contribution compared to the linear term $D_{L1} L^-/q^-$ if the longitudinal drag (momentum loss) is small or the energy of the hard parton is very high. 
This indicates that for small longitudinal momentum loss and only considering the leading contribution from the drag and diffusions of the hard parton's momentum, the medium-induced photon bremsstrahlung spectra are the same for single and multiple scattering scenarios.
We note that the above formula for medium-induced single photon bremsstrahlung spectrum can be directly used as the input to the phenomenological study of jet-medium photons in relativistic heavy-ion collisions, which are expected to give significant contribution to direct photon production in the intermediate transverse momentum regime \cite{Fries:2002kt, Qin:2009bk}.

\section{Summary}

Within the framework of deep-inelastic scattering off a large nucleus, we have studied the single photon bremsstrahlung process from a hard quark jet through the scattering with the constituents of a dense medium.
We have included the effects from both longitudinal and transverse momentum exchanges between the hard parton and the medium constituents on the medium-induced photon emission process.
Using a gradient expansion for the exchanged momentum up to the second order, we have derived a closed form for the single photon radiation spectrum with the incorporation of the contributions from the transverse momentum diffusion as well as the longitudinal momentum drag and diffusion of the propagating jet parton.
It is found that while the transverse momentum broadening on the propagating jet parton induces additional photon radiation in the medium, the longitudinal drag tends to suppress the medium-induced photon emission.
By comparing the results from single scattering and from the resummation of multiple scatterings, we find an additional term for the case of multiple scatterings due to the coupling between different scatterings experienced by the propagating hard jet parton.
Since the additional term is quadratic in the drag coefficient, it is suppressed compared to the leading contribution when the momentum transfer between jet and medium is small, i.e., the photon radiation spectra for two single and multiple scattering scenarios are the same if the leading contributions from the drag and the diffusions of the hard parton's momentum are considered. 
The medium-induced photon radiation spectrum obtained in this work can be directly applied to the phenomenological studies of the production of jet-medium photons in heavy-ion collisions.
The study of medium-induced photon emission in this work also serves as an intermediate step for the investigation of the medium-induced gluon emission from a quark jet interacting with a dense nuclear medium.
We leave them for future effort.

\section*{Acknowledgments}

This work is supported in part by Natural Science Foundation of China (NSFC) under grant Nos. 11375072, 113575070, 11221504, 11135011, and by Ministry of Science and Technology of China (MSTC) under ``973" project No. 2015CB856904(4).

\begin{widetext}
\section*{Appendix}

First, we provide the main results for the diagrams with central cuts mentioned in Sec. III, i.e., a single scattering in both the amplitude and the complex conjugate, as shown in Fig. \ref{11twist}.
The hard part of the matrix element for Fig. \ref{11twist}(a) reads as:
\begin{eqnarray}
\label{linear_approximation}
T_{(a)} &=& \frac{2yP(y)}{l_\perp^2}.
\end{eqnarray}
The phase factor for Fig. \ref{11twist}(a) can be simplified as:
\begin{eqnarray}
S_{(a)}&=&e^{-ix_{D0}p^+y_0^-} e^{-i x_{C1} p^+ y_1^-}e^{ix_{C1}' p^+ {y}_1'^-}\left(e^{-i\delta x_{D0} p^+ y_0^-} - e^{-i\delta x_{D0} p^+ y_1^-}\right) \left( 1 - e^{i\delta x_{D0}' p^+ {y}_1'^-}\right)
\nonumber\\
&\approx& e^{-ix_Lp^+y_0^-} [2-2 \cos(x_L p^+ Y_1^-)].
\end{eqnarray}
The hadronic tensor for the Fig. \ref{11twist}(a) reads as:
\begin{eqnarray}
W^{A\mu\nu}_{(a)}
\!\!&\approx&\!\!
(-g_\perp^{\mu\nu}) A C_p^A \sum_q Q_q^4 \frac{\alpha_e}{2\pi}
\int dy P(y) \int \frac{d^2l_\perp}{\pi l_\perp^2} \int{d^3\mathbf{l}_q}
\int dy_0^- e^{-i(x_B+x_L)p^+y_0^-}  \langle p | \bar{\psi}(y_0^-) \frac{\gamma^+}{2} \psi(0) |p\rangle
\nonumber\\ \!\!&\times&\!\!
\int d Y_1^- \int d \delta y_1^- \left(g^2 \frac{C_F}{N_c^2-1}\frac{\rho}{2 p^+}\right) \int d^3\delta \mathbf{y}_1 \int \frac{d^3\mathbf{p}_1}{(2\pi)^3} e^{-i\mathbf{p}_1\cdot \delta \mathbf{y}_1} \langle p| A^+(\delta{y_1}) A^+(0) | p\rangle
\nonumber\\ \!\!&\times&\!\!
[2-2 \cos(x_L p^+ Y_1^-)]
\delta^3(\mathbf{l}+\mathbf{l}_q - \mathbf{p}_1 - \mathbf{q}).
\end{eqnarray}
The hard part of the matrix element for Fig. \ref{11twist}(b) reads as:
\begin{eqnarray}
T_{(b)} &\approx& \frac{2yP(y)}{l_\perp^2} \left[1 +  \frac{2y(1-y)}{1+(1-y)^2} \frac{p_1^-}{q^-} + 2 y \frac{\mathbf{l}_\perp \cdot \mathbf{p}_{1\perp}}{l_\perp^2}- \frac{y (2 - 3y)}{1+(1-y)^2}\left(\frac{p_1^-}{q^-}\right)^2 -y^2\frac{p_{1\perp}^2}{l_\perp^2}  +4y^2\frac{(\mathbf{l}_\perp \cdot \mathbf{p}_{1\perp})^2}{l_\perp^4}
\nonumber\right.
\\ &&
\left.+ 2y \left(\frac{2y(1-y)}{1+(1-y)^2}-1\right) \frac{p_1^-}{q^-} \frac{\mathbf{l}_\perp \cdot \mathbf{p}_{1\perp}}{l_\perp^2}\right].
\end{eqnarray}
The phase factor for Fig. \ref{11twist}(b) can be simplified as:
\begin{eqnarray}
S_{(b)}&=&e^{-ix_{D0}p^+y_0^-}e^{-ix_{D1}p^+y_1^-}e^{i{x}_{D1}'p^+{y}_1'^-}\left( e^{-i\delta \bar{x}_{D1} p^+ y_1^-} e^{i\delta \bar{x}_{D1}' p^+ {y}_1'^-}\right)
\nonumber\\
&\approx& e^{-ix_Lp^+y_0^-}.
\end{eqnarray}
The hadronic tensor for the Fig. \ref{11twist}(b) reads as:
\begin{eqnarray}
W^{A\mu\nu}_{(b)}
\!\!&\approx&\!\!
(-g_\perp^{\mu\nu}) A C_p^A \sum_q Q_q^4 \frac{\alpha_e}{2\pi}
\int dy P(y) \int \frac{d^2l_\perp}{\pi l_\perp^2} \int{d^3\mathbf{l}_q}
\int dy_0^- e^{-i(x_B+x_L)p^+y_0^-}  \langle p | \bar{\psi}(y_0^-) \frac{\gamma^+}{2} \psi(0) |p\rangle
\nonumber\\ \!\!&\times&\!\!
\int d Y_1^- \int d \delta y_1^- \left(g^2 \frac{C_F}{N_c^2-1}\frac{\rho}{2 p^+}\right) \int d^3\delta \mathbf{y}_1 \int \frac{d^3\mathbf{p}_1}{(2\pi)^3} e^{-i\mathbf{p}_1\cdot \delta \mathbf{y}_1} \langle p| A^+(\delta{y_1}) A^+(0) | p\rangle
\nonumber\\ \!\!&\times&\!\!
\left[1 +  \frac{2y(1-y)}{1+(1-y)^2} \frac{p_1^-}{q^-} + 2 y \frac{\mathbf{l}_\perp \cdot \mathbf{p}_{1\perp}}{l_\perp^2}- \frac{y (2 - 3y)}{1+(1-y)^2}\left(\frac{p_1^-}{q^-}\right)^2 -y^2\frac{p_{1\perp}^2}{l_\perp^2}  +4y^2\frac{(\mathbf{l}_\perp \cdot \mathbf{p}_{1\perp})^2}{l_\perp^4}
\nonumber\right.
\\ &&
\left.+ 2y \left(\frac{2y(1-y)}{1+(1-y)^2}-1\right) \frac{p_1^-}{q^-} \frac{\mathbf{l}_\perp \cdot \mathbf{p}_{1\perp}}{l_\perp^2}\right]
\delta^3(\mathbf{l}+\mathbf{l}_q - \mathbf{p}_1 - \mathbf{q}).
\end{eqnarray}
The hard part of the matrix element for Fig. \ref{11twist}(c) reads as:
\begin{eqnarray}
T_{(c)} &\approx& \frac{2yP(y)}{l_\perp^2}
\left[1 + \frac{y(1-y)}{1+(1-y)^2} \frac{p_1^-}{q^-} +  y \frac{\mathbf{l}_\perp \cdot \mathbf{p}_{1\perp}}{l_\perp^2} - \frac{y (1-y)}{1+(1-y)^2} \left(\frac{p_1^-}{q^-}\right)^2-y^2\frac{p_{1\perp}^2}{l_\perp^2} + 2y^2\frac{(\mathbf{l}_\perp \cdot \mathbf{p}_{1\perp})^2}{l_\perp^4}
\nonumber\right. \\ &&
\left.
+ y \left(\frac{y(1-y)}{1+(1-y)^2}-1\right) \frac{p_1^-}{q^-} \frac{\mathbf{l}_\perp \cdot \mathbf{p}_{1\perp}}{l_\perp^2}
\right].
\end{eqnarray}
The phase factor for Fig. \ref{11twist}(c) can be simplified as:
\begin{eqnarray}
S_{(c)}&=&e^{-ix_{D0}p^+y_0^-}e^{-ix_{D1}p^+y_1^-}e^{i{x}_{C1}'p^+{y}_1'^-} e^{-i\delta \bar{x}_{D1} p^+ y_1^-}\left( 1 - e^{i\delta x_{D0}' p^+ {y}_1'^-}\right)
\nonumber\\
&\approx& e^{-ix_Lp^+y_0^-}(e^{-i x_L p^+ Y_1^-}-1);
\end{eqnarray}
Fig. \ref{11twist}(d) is the complex conjugate of Fig. \ref{11twist}(c).
The hard part of the matrix element for Fig. \ref{11twist}(d) reads as:
\begin{eqnarray}
T_{(d)} &=& T_{(c)}.
\end{eqnarray}
The phase factor for Fig. \ref{11twist}(d) can be simplified as:
\begin{eqnarray}
S_{(d)}&=& e^{-ix_{D0}p^+y_0^-}e^{-ix_{C1}p^+y_1^-}e^{i{x}_{D1}'p^+{y}_1'^-}\left(e^{-i\delta x_{D0} p^+ y_0^-} - e^{-i\delta x_{D0} p^+ y_1^-}\right) e^{i\delta \bar{x}_{D1}' p^+ {y}_1'^-}
\nonumber\\
&\approx& e^{-ix_Lp^+y_0^-}(e^{i x_L p^+ Y_1^-}-1).
\end{eqnarray}
The hadronic tensors for the Fig. \ref{11twist}(c) and Fig. \ref{11twist}(d) read as:
\begin{eqnarray}
W^{A\mu\nu}_{(c)} +  W^{A\mu\nu}_{(d)}
\!\!&\approx&\!\!
(-g_\perp^{\mu\nu}) A C_p^A \sum_q Q_q^4 \frac{\alpha_e}{2\pi}
\int dy P(y) \int \frac{d^2l_\perp}{\pi l_\perp^2} \int{d^3\mathbf{l}_q}
\int dy_0^- e^{-i(x_B+x_L)p^+y_0^-}  \langle p | \bar{\psi}(y_0^-) \frac{\gamma^+}{2} \psi(0) |p\rangle
\nonumber\\ \!\!&\times&\!\!
\int d Y_1^- \int d \delta y_1^- \left(g^2 \frac{C_F}{N_c^2-1}\frac{\rho}{2 p^+}\right) \int d^3\delta \mathbf{y}_1 \int \frac{d^3\mathbf{p}_1}{(2\pi)^3} e^{-i\mathbf{p}_1\cdot \delta \mathbf{y}_1} \langle p| A^+(\delta{y_1}) A^+(0) | p\rangle
\nonumber\\ \!\!&\times&\!\!
\left[1 + \frac{y(1-y)}{1+(1-y)^2} \frac{p_1^-}{q^-} +  y \frac{\mathbf{l}_\perp \cdot \mathbf{p}_{1\perp}}{l_\perp^2} - \frac{y (1-y)}{1+(1-y)^2} \left(\frac{p_1^-}{q^-}\right)^2-y^2\frac{p_{1\perp}^2}{l_\perp^2} + 2y^2\frac{(\mathbf{l}_\perp \cdot \mathbf{p}_{1\perp})^2}{l_\perp^4}
\nonumber\right. \\ &&
\left.
+ y \left(\frac{y(1-y)}{1+(1-y)^2}-1\right) \frac{p_1^-}{q^-} \frac{\mathbf{l}_\perp \cdot \mathbf{p}_{1\perp}}{l_\perp^2}
\right]
\left[2 \cos(x_L p^+ Y_1^-) - 2\right]
\delta^3(\mathbf{l}+\mathbf{l}_q - \mathbf{p}_1 - \mathbf{q}).
\end{eqnarray}

Now, we provide the main results for the diagrams with non-central cuts mentioned in Sec. III, i.e., two gluon insertions in the amplitude and no gluon insertion in the complex conjugate (or vice versa), as shown in Fig. \ref{20twist}.
The hard part of the matrix element for Fig. \ref{20twist}(e) reads as:
\begin{eqnarray}
T_{(e)} = \frac{2yP(y)}{l_\perp^2}
\end{eqnarray}
The phase factor for Fig. \ref{20twist}(e) can be simplified as:
\begin{eqnarray}
S_{(e)} &=& e^{-i x_{D0} p^+ y_0^-}e^{i x_{C1}' p^+ y_1'^-}e^{i x_{C2}' p^+ y_2'^-}e^{-i \delta x_{D0} p^+ y_0^-}(1-e^{i \delta x_{D0}' p^+ y_1'^-})
\nonumber\\&\approx& e^{-i x_L p^+ y_0^-}(1-e^{i x_L p^+ Y_1^-})
\end{eqnarray}
Fig. \ref{20twist}(f) is the complex conjugate of Fig. \ref{20twist}(e).
The hard part of the matrix element for Fig. \ref{20twist}(f) reads as:
\begin{eqnarray}
T_{(f)}=T_{(e)}
\end{eqnarray}
The phase factor for Fig. \ref{20twist}(f) can be simplified as:
\begin{eqnarray}
S_{(f)}&=& e^{-i x_{D0}p^+ y_0^-} e^{-i x_{C1} p^+ y_1^-}e^{-i x_{C2} p^+ y_2^-}(e^{-i \delta x_{D0} p^+ y_0^-}-e^{-i \delta x_{D0} p^+ y_1^-})
\nonumber\\&\approx& e^{-i x_L p^+ y_0^-}(1-e^{-i x_L p^+ Y_1^-})
\end{eqnarray}
The hadronic tensors for the Fig. \ref{20twist}(e) and Fig. \ref{20twist}(f) read as:
\begin{eqnarray}
\label{aa}
W^{A\mu\nu}_{(e)} + W^{A\mu\nu}_{(f)}
\!\!&\approx&\!\!
(-g_\perp^{\mu\nu}) A C_p^A\sum_q Q_q^4 \frac{\alpha_e}{2\pi} \int dy P(y) \int \frac{d^2l_{\perp}}{\pi l^2_{\perp}}
\int d^3\mathbf{l}_q  (2\pi) f_q(x_B+x_L)
\delta^3(\mathbf{l}+\mathbf{l}_q - \mathbf{q}).
\nonumber\\\!\!&\times&\!\!
\int d {Y_1}^- \left[-1 + \cos(x_L p^+ {Y_1}^-)\right]
\int d \delta {y_1}^-  \left(g^2 \frac{C_F}{N^2_c-1} \frac{\rho}{2 p^+}\right) \langle p | A^+(\delta {y_1}^-) A^+(0)  |p\rangle
\end{eqnarray}
The hard part of the matrix element for Fig. \ref{20twist}(g) reads as:
\begin{eqnarray}
T_{(g)}=\frac{2yP(y)}{l_\perp^2}
\end{eqnarray}
The phase factor for Fig. \ref{20twist}(g) can be simplified as:
\begin{eqnarray}
S_{(g)}& =&e^{-i x_{D0} p^+ y_0^-}e^{i x_{D1}' p^+ y_1'^-}e^{i x_{D2}' p^+ y_2'^-} e^{-i \delta {x}_{D0} p^+ y_0^-} e^{i \delta \bar{x}_{D2}' p^+ y_2'^-}
\nonumber\\&\approx& e^{-i x_L p^+ y_0^-} e^{i x_L p^+ Y_1^-}
\end{eqnarray}
Fig. \ref{20twist}(h) is the complex conjugate of Fig. \ref{20twist}(g).
The hard part of the matrix element for Fig. \ref{20twist}(h) reads as:
\begin{eqnarray}
T_{(h)}= T_{(g)}
\end{eqnarray}
The phase factor for Fig. \ref{20twist}(h) can be simplified as:
\begin{eqnarray}
S_{(h)} &=&e^{-i x_{D0} p^+ y_0^-}e^{-i x_{D1} p^+ y_1^-}e^{-i x_{D2} p^+ y_2^-} e^{-i \delta \bar{x}_{D2}p^+ y_2^-}
\nonumber\\&\approx& e^{-i x_L p^+ y_0^-} e^{-i x_L p^+ Y_1^-}
\end{eqnarray}
The hadronic tensors for the Fig. \ref{20twist}(g) and Fig. \ref{20twist}(h) read as:
\begin{eqnarray}
W^{A\mu\nu}_{(g)} + W^{A\mu\nu}_{(h)}
\!\!&\approx&\!\!
(-g_\perp^{\mu\nu}) A C_p^A\sum_q Q_q^4 \frac{\alpha_e}{2\pi} \int dy P(y) \int \frac{d^2l_{\perp}}{\pi l^2_{\perp}}
\int d^3\mathbf{l}_q  (2\pi) f_q(x_B+x_L)
\delta^3(\mathbf{l}+\mathbf{l}_q - \mathbf{q}).
\nonumber\\\!\!&\times&\!\!
\int d {Y_1}^- \left[- \cos(x_L p^+ {Y_1}^-)\right] \int d \delta {y_1}^-  \left(g^2 \frac{C_F}{N^2_c-1} \frac{\rho}{2 p^+}\right) \langle p | A^+(\delta {y}_1^-) A^+(0)  |p\rangle
\end{eqnarray}
The hard part of the matrix element for Fig. \ref{20twist}(i) reads as:
\begin{eqnarray}
T_{(i)}&\approx& \frac{2yP(y)}{l_\perp^2}
\left[1 + \frac{y(1-y)}{1+(1-y)^2} \frac{p_1^-}{q^-} +  y \frac{\mathbf{l}_\perp \cdot \mathbf{p}_{1\perp}}{l_\perp^2} - \frac{y (1-y)}{1+(1-y)^2} \frac{{p_1^-}^2}{{q^-}^2}-y^2\frac{p_{1\perp}^2}{l_\perp^2} + 2y^2\frac{(\mathbf{l}_\perp \cdot \mathbf{p}_{1\perp})^2}{l_\perp^4}
\nonumber\right. \\ &&
\left.
+ y \left(\frac{y(1-y)}{1+(1-y)^2}-1\right) \frac{p_1^-}{q^-} \frac{\mathbf{l}_\perp \cdot \mathbf{p}_{1\perp}}{l_\perp^2}
\right]
\end{eqnarray}
The phase factor for Fig. \ref{20twist}(i) can be simplified as:
\begin{eqnarray}
S_{(i)}&=&e^{- x_{D0} p^+ y_0^-} e^{i x'_{D1} p^+ y_1'^-} e^{i x_{C2}' p^+ y_2'^-} e^{-i \delta {x}_{D0} p^+ y_0^-} \left(e^{i \delta \bar{x}_{D1}' p^+ y_1'^-} - e^{i \delta \bar{x}_{D1}' p^+ {y'_2}^-}\right)
\nonumber\\&\approx& e^{-i x_L p^+ y_0^-}\left(e^{i x_L p^+ (Y_1^- + \frac{1}{2} \delta y_1^-)} - e^{i x_L p^+ (Y_1^- - \frac{1}{2} \delta y_1^-)}\right)
\approx 0
\end{eqnarray}
Fig. \ref{20twist}(j) is the complex conjugate of Fig. \ref{20twist}(i).
The hard part of the matrix element for Fig. \ref{20twist}(i) reads as:
\begin{eqnarray}
T_{(j)}&=& T_{(i)}
\end{eqnarray}
The phase factor for Fig. \ref{20twist}(j) can be simplified as:
\begin{eqnarray}
S_{(j)}&=& e^{-i x_{D0} p^+ y_0^-} e^{-i x_{D1} p^+ y_1^-} e^{-i x_{C2} p^+ y_2^-} \left(e^{i \delta \bar{x}_{D1} p^+ y_1^-} - e^{i \delta \bar{x}_{D1} p^+ y_2^-}\right)
\nonumber\\&\approx& e^{-i x_L p^+ y_0^-}\left(e^{i x_L p^+ (Y_1^- + \frac{1}{2} \delta y_1^-)} - e^{i x_L p^+ (Y_1^- - \frac{1}{2} \delta y_1^-)}\right)
\approx 0
\end{eqnarray}
Since the phase factors for Fig. \ref{20twist}(i) and Fig. \ref{20twist}(j) are approximately zero, they provide vanishing contributions to the hadronic tensors:
\begin{eqnarray}
W^{A\mu\nu}_{(i)} + W^{A\mu\nu}_{(j)} \!\!&\approx&\!\! 0
\end{eqnarray}
\end{widetext}


\bibliographystyle{h-physrev5}
\bibliography{GYQ_refs}
\end{document}